\documentclass[useAMS,usenatbib]{mn2e}
\usepackage{graphicx}

\title[SDSS Clustering Evolution]{Evolution of the Clustering of Photometrically Selected SDSS Galaxies}

\author[A. J. Ross, W. J. Percival, \& R. J. Brunner]{Ashley J. Ross\thanks{Email: Ashley.Ross@port.ac.uk}$^{1,2}$,Will J. Percival$^{1}$, \&  Robert J. Brunner$^{2,3}$ \\
$^1$Institute of Cosmology \& Gravitation, Dennis Sciama Building, University of Portsmouth, Portsmouth, PO1 3FX, UK\\
$^2$Department of Astronomy, University of Illinois, 1002 W Green St., Urbana, IL 61801, USA\\
$^3$National Center for Supercomputing Applications, Champaign, IL 61820, USA}

\begin{document}

\date{Accepted for publication in MNRAS}

\pagerange{\pageref{firstpage}--\pageref{lastpage}} \pubyear{2010}

\maketitle

\label{firstpage}

\begin{abstract}
We measure the angular auto-correlation functions, $\omega(\theta)$, of SDSS galaxies selected to have photometric redshifts $0.1 < z < 0.4$ and absolute $r$-band magnitudes $M_r < -21.2$.  We split these galaxies into five overlapping redshift shells of width 0.1 and measure $\omega(\theta)$ in each subsample in order to investigate the evolution of SDSS galaxies.  We find that the bias increases substantially with redshift --- much more so than one would expect for a passively evolving sample.  We use halo-model analysis to determine the best-fit halo-occupation-distribution (HOD) for each subsample, and the best-fit models allow us to interpret the change in bias physically.  In order to properly interpret our best-fit HODs, we convert each halo mass to its $z = 0$ passively evolved bias ($b_{\rm o}$), enabling a direct comparison of the best-fit HODs at different redshifts.  We find that the minimum halo $b_{\rm o}$ required to host a galaxy decreases as the redshift decreases, suggesting that galaxies with $M_r < -21.2$ are forming in halos at the low-mass end of the HODs over our redshift range.  We use the best-fit HODs to determine the change in occupation number divided by the change in mass of halos with constant $b_{\rm o}$, $\Delta N/\Delta M (b_{\rm o})$, and we find a sharp peak at $b_{\rm o} \sim 0.9$ --- corresponding to an average halo mass of $\sim 10^{12}h^{-1}M_{\odot}$.  We thus present the following scenario:  the bias of galaxies with $M_r < -21.2$ decreases as the Universe evolves because these galaxies form in halos of mass $\sim 10^{12}h^{-1}M_{\odot}$ (independent of redshift), and the bias of these halos naturally decreases as the Universe evolves.
\end{abstract}


\section{Introduction}
The angular auto-correlation function of galaxies, $\omega(\theta)$, encodes a wealth of information, about both cosmology and the properties of galaxies.  Wide-field surveys, such as the Sloan Digital Sky Survey (SDSS), allow precise calculations of $\omega(\theta)$ over a range of scales spanning over three orders of magnitude --- thereby probing both the clustering of galaxies dominated by interactions within dark matter halos and also the clustering of galaxies that is determined by the matter density field.  Angular clustering measurements made using data from photometric surveys are complicated by the fact that such surveys can only easily provide precise information on the locations of galaxies in two dimensions, while many analyses of interest require knowledge of the three dimensional distribution.  Multi-band surveys, such as SDSS, allow estimation of photometric redshifts, and thus with careful treatment, one can estimate the radial distribution of galaxies and split the galaxies by redshift, type, and luminosity.  As a result, one can investigate the evolution of galaxies with photometric data.  The techniques to do this are gaining in importance, as many of the next generation of wide-field surveys will rely primarily on photometric redshifts to gain knowledge of their respective radial distributions (e.g., DES, PanStarrs, LSST).  

Using the auto-correlation function of galaxies to study their properties has been aided in recent years by the development of the `halo-model' (see, e.g. \citealt{Kauf97,PS00,CooSh02,zheng05,Tinker}) as a way of parameterising galaxy bias.  One can fill dark matter halos with galaxies based on a statistical `halo-occupation-distribution' (HOD), allowing one to model the clustering of galaxies within halos (and thus non-linear scales) while providing a self consistent determination of the bias at linear scales.  Thus, as shown by, e.g., \cite{Z04}, \cite{Blake}, \cite{Tink08}, Ross \& Brunner (2009; hereafter R09) one can use measurements of galaxy auto-correlation functions to constrain the HODs of different sets of galaxies and to gain information on the nature in which galaxies occupy dark matter halos.  

Many recent studies have used clustering measurements to study the evolution of galaxies.  \cite{Wake08} and \cite{Brown08} measured the auto-correlation functions of luminous red galaxies (LRGs) and interpreted the results with the halo model to show that their evolution is inconsistent with passive evolution, while \cite{To10} were able to determine, via luminosity weighted power-spectrum measurements, that the non-passive evolution is due primarily to lower luminosity LRGs.  \cite{ZZ07} used auto-correlation function measurements to constrain the HODs of SDSS spectroscopic galaxies ($z \sim 0.1$) and DEEP2 galaxies ($z \sim 1$), allowing them to investigate the evolution of the HOD, stellar mass, and satellite fraction of galaxies over a relatively large range of luminosities.  Ross et al. (2007; hereafter R07) studied the clustering of SDSS DR5 galaxies split by redshift between $z < 0.3$ and $0.3 < z < 0.4$, and found significantly larger bias at higher redshift, especially for late-type galaxies.

In this paper, we use galaxies photometrically selected from the SDSS seventh data release (DR7) to investigate the evolution of galaxies between redshifts 0.1 and 0.4.  While this represents a relatively small range in redshift, our study offers a unique combination in that it utilizes over 6000 square degrees of observing area (after masking) and the evolution we study is for galaxies drawn entirely from the SDSS (and we thus do not have to worry about selection techniques of different surveys or differing calibration issues).  We are thus able to precisely measure the angular auto-correlations of SDSS galaxies, use the halo-model to interpret them, and self-consistently compare the results at different redshifts.

Our paper is outlined as follows:  \S\ref{sec:data} describes the creation of our galaxy catalog and its five subsamples, its angular masking, and our methods for estimating the redshift distributions of each subsample; \S\ref{sec:tools} describes how we measure the angular auto-correlation functions of these galaxies and how we model the results;  \S\ref{sec:res} presents the results of our auto-correlation function measurements and the best-fit HOD for each redshift slice; in \S5 we use cross-correlation measurements to investigate potential systematics, in \S\ref{sec:phys} we discuss the physical implications of our results; finally, we conclude in \S\ref{sec:con}.  Throughout this work, we assume a flat cosmology with $\Omega_m = 0.3$, $h = 0.7$, $\sigma_8 = 0.8$, and $\Omega_b = 0.05$.

\section{Data}
\label{sec:data}
We use data from the Northern, contiguous area of the SDSS seventh data release (DR7).  This survey obtains wide-field CCD photometry (\citealt{C}) in five passbands ($u,g,r,i,z$; e.g., \citealt{F}).  DR7 contains a moderate increase over the DR5 imaging area  ($\sim$ 500 square degrees), but the precision and accuracy of its photometric redshift catalog represents a substantial improvement over previous data releases (\citealt{DR7}).  We select galaxies from the DR7 {\rm photoz} catalog with de-reddened $r$-band magnitudes ($r_d$) less than 21.  Redshifts in this catalog were estimated using a hybrid template/empirical  approach, and the output includes rest-frame absolute magnitudes, k-corrections, and galaxy-type values.  We use this information to construct an approximately volume-limited sample, using the methods outlined in \cite{Bud03}.  Our resulting criteria are that galaxies have photometric redshifts $z < 0.4$ and $r$-band absolute magnitudes, $M_r  < -21.2$ (equivalent to $M_r - 5{\rm log}_{10} h < -20.43 $).  

\begin{figure}
\includegraphics[width=84mm]{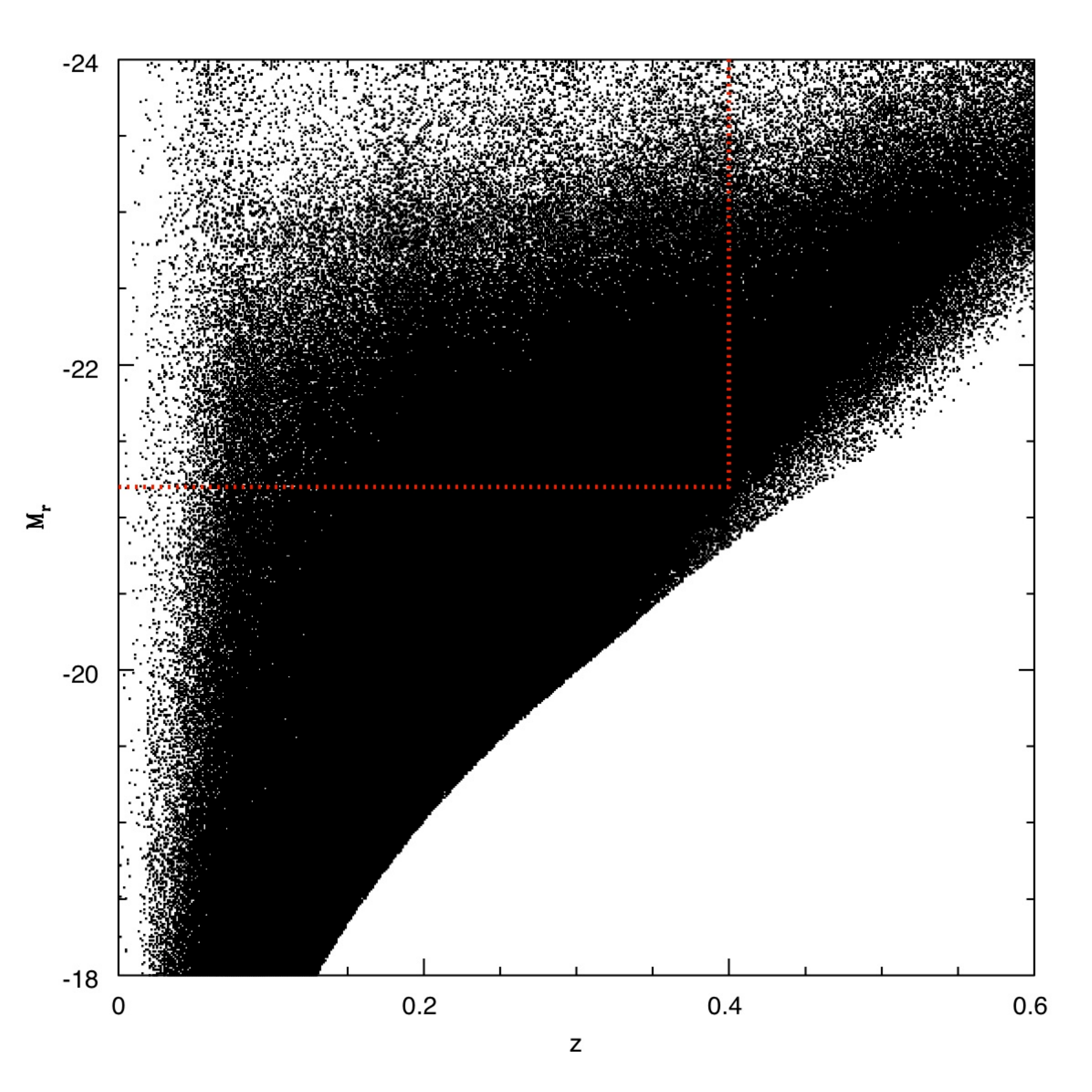}
\caption{The $M_r, z$ parameter-space in which we find galaxies is shaded black.  Red dotted lines define the boundaries of our selection criteria ($M_r < -21.2$, $z < 0.4$).}
\label{fig:mvz}
\end{figure}

Figure \ref{fig:mvz} displays our galaxy selection criteria in dotted redlines, plotted over the shaded black region of $M_r , z$ parameter space where galaxies with $r_d < 21$ exist in the SDSS DR7 {\rm photoz} catalog.  We make our selection at $M_r < -21.2$, rather than the very edge of the shaded region ($\sim -20.7$), in order to account for differences in k-corrections between different galaxy types (which can be as high as 1 at $z = 0.4$ in the $r$-band).  We applied imaging masks and cuts on seeing and reddening at $1\farcs5$ and $A_r = 0.2$ (as in R07) to the survey area, while cutting out data with flags indicating poor photometry/spurious object detection.  This left a total of 3,123,487 galaxies with $0.1 < z < 0.4$, occupying 6019 square degrees of observed sky.

We split the sample by redshift into five samples with $0.1 < z < 0.2$, $0.15 < z < 0.25$, $0.2 < z < 0.3$, $0.25 < z < 0.35$, and $0.3 < z < 0.4$.  While splitting the samples in this manner means that they are not mutually exclusive, it allows for a test of whether the redshift evolution is smooth (any sharp transition might imply a systematic in the data).  Reducing the width of the redshift slices further would not provide significantly more information, as the error on the photometric redshifts is $\sim 0.05$ at $z \sim 0.3$.  

\subsection{Estimating True Redshift Distributions}
We take care in estimating the form of each of our redshift distributions, as this is quite important to our analysis.  We treat each individual galaxy's redshift as a Gaussian PDF based on its maximum likelihood redshift and associated error, and convolve this PDF with volume and luminosity function (LF) constraints.  Volume arguments imply that a galaxy is more likely to have a larger redshift than a smaller one, while LF arguments imply that a galaxy is more likely to be faint than bright.  Thus, we sample the Gaussian PDF to find a redshift we refer to as $z^{\prime}$ and determine the sampled absolute magnitude, $M^{\prime}$, by adding to $M_r$ the difference in distance modulus between $z$ and $z^{\prime}$.  In order to apply the volume and LF constraints, we weight each sampled redshift by 
\begin{equation}
f_{nz} = (x(z^{\prime})/x(z))^2\Phi(M^{\prime})/\Phi(M_{r}),
\end{equation}
where $x(z)$ is the comoving distance to redshift $z$, and $\Phi(M)$ is the best-fit Schecter form of the LF determined by \cite{DR6LF} for $r$-band SDSS galaxies.  

We sample each galaxy's Gaussian PDF 10 times and find $f_{nz}$ for each sampling.  We normalise such that the sum of $f_{nz}$ adds to 1 for each galaxy (in order to insure that each galaxy contributes to the overall $dN/dz$ at the same level) and then add each of the 10 normalised $f_{nz}$ to their appropriate bin (we use bins of width $\Delta z = 0.001$).  Thus, when completed for galaxies in a given sample, we have an estimate for the total number galaxies at $z \pm 0.0005$ included in the sample.  Each $dN/dz$ is then normalised by dividing each bin by $\sum_i dN/dz(z_i)*0.001$ and we interpolate between bins to obtain a continuous, normalised, $dN/dz$.  

Our construction of $dN/dz$ eliminates unphysical results --- such as non-zero probability at redshift 0.  It is similar to the treatment presented in section 4.2 of \cite{Bud03}, but we apply the treatment to each galaxy rather than bin in magnitude.  In general, the resulting redshift distributions are similar to the distributions one gets from Gaussian sampling (the LF and volume effects tend to cancel each other) but have lower values at the tails of the distribution.  

\begin{figure}
\includegraphics[width=84mm]{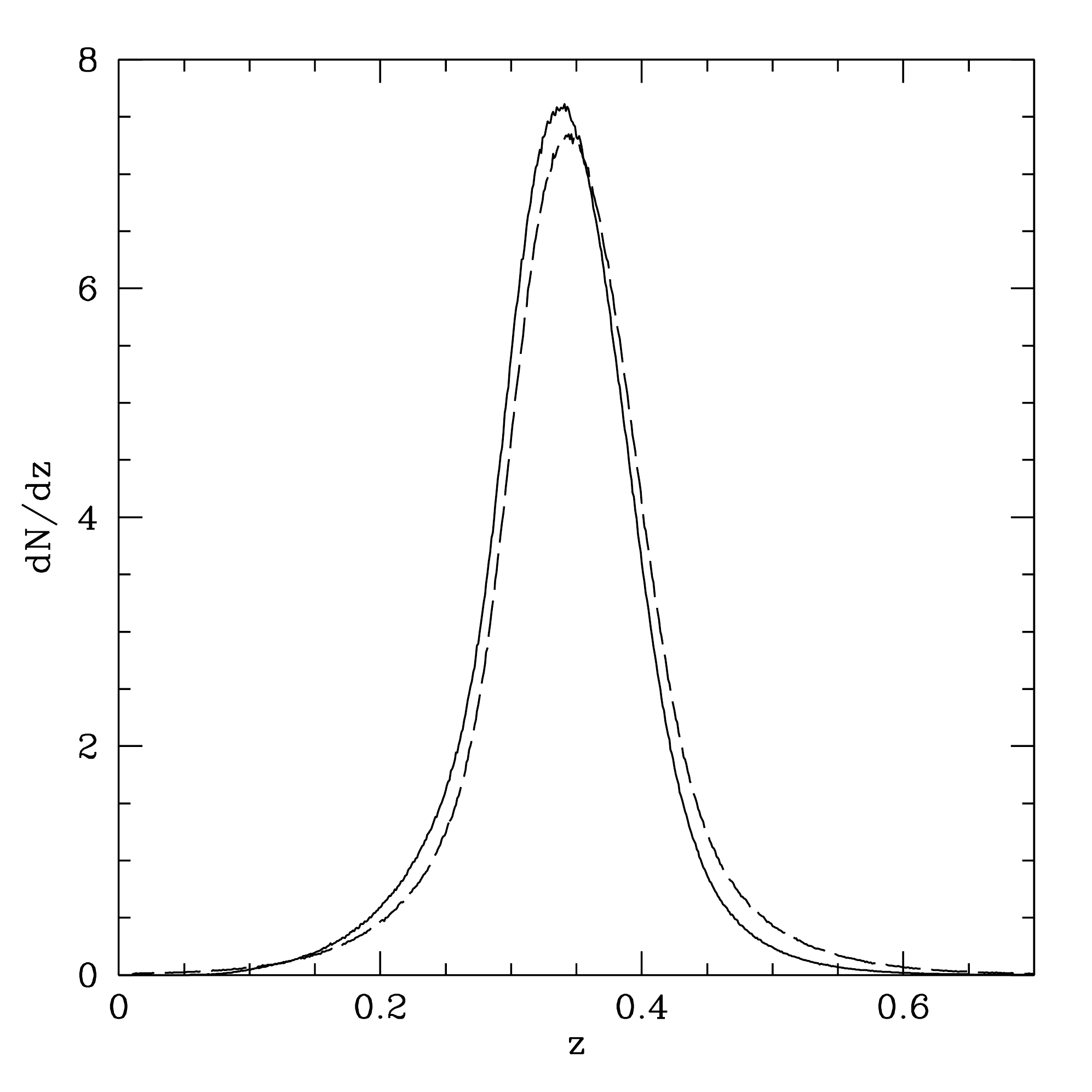}
\caption{The normalised (such that they integrate to 1) redshift distributions for galaxies in our sample with $0.3 < z < 0.4$, determined combining Gaussian sampling with LF and volume considerations (solid line) and using only Gaussian sampling (dashed line).}
\label{fig:nzcom}
\end{figure}

Figure \ref{fig:nzcom} displays the normalised (such that they integrate to 1) redshift distributions of our galaxy sample with $0.3 < z < 0.4$. The solid line displays $dN/dz$ determined using Gaussian sampling combined with LF and volume considerations, while the dashed line displays the result obtained using only Gaussian sampling.  For this sample, the median redshift is shifted to a slightly lower value, and the distribution has lower values at the tails (resulting in a stronger peak).  The shift in redshift is due to the fact that our galaxies have magnitudes close to $M_*$, and the LF thus suggests that a decrease in redshift is more likely than an increase since the decrease will lower the luminosity.  

\begin{figure}
\includegraphics[width=84mm]{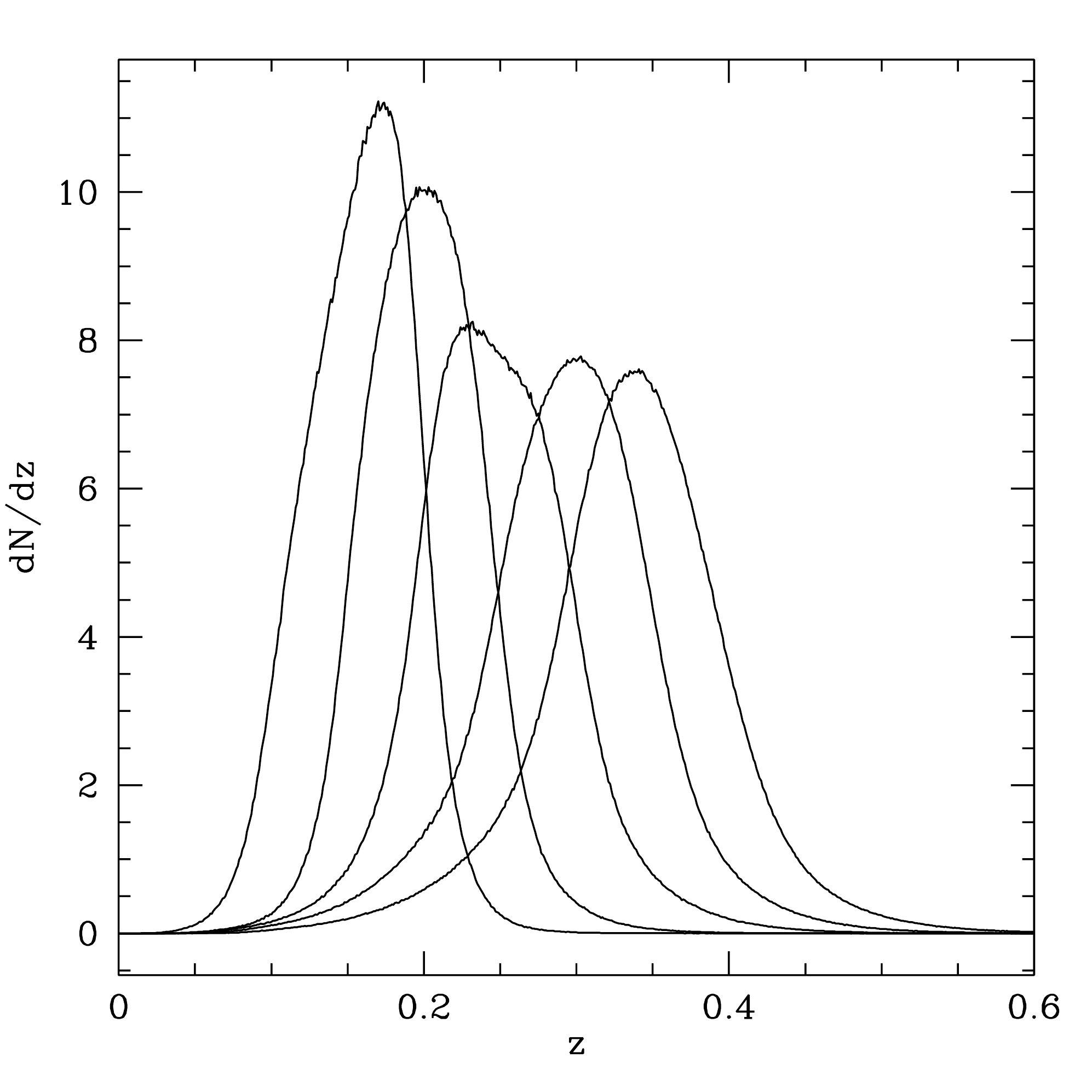}
\caption{The normalised (such that they integrate to 1) redshift distributions for each of the five photometric redshift slices $0.1 < z < 0.2$, $0.15 < z < 0.25$, $0.2 < z < 0.3$, $0.25 < z < 0.35$, and $0.3 < z < 0.4$.}
\label{fig:nzDR7}
\end{figure}

Figure \ref{fig:nzDR7} displays the normalised redshift distributions for each redshift slice (incorporating the LF and volume effects, as one can assume we do from here-on).  The distributions get wider as the median redshift increases due to the fact that the mean redshift error increases with redshift (see Table \ref{tab:DR7z4}).  Each individual distribution appears roughly Gaussian.

\begin{figure}
\includegraphics[width=84mm]{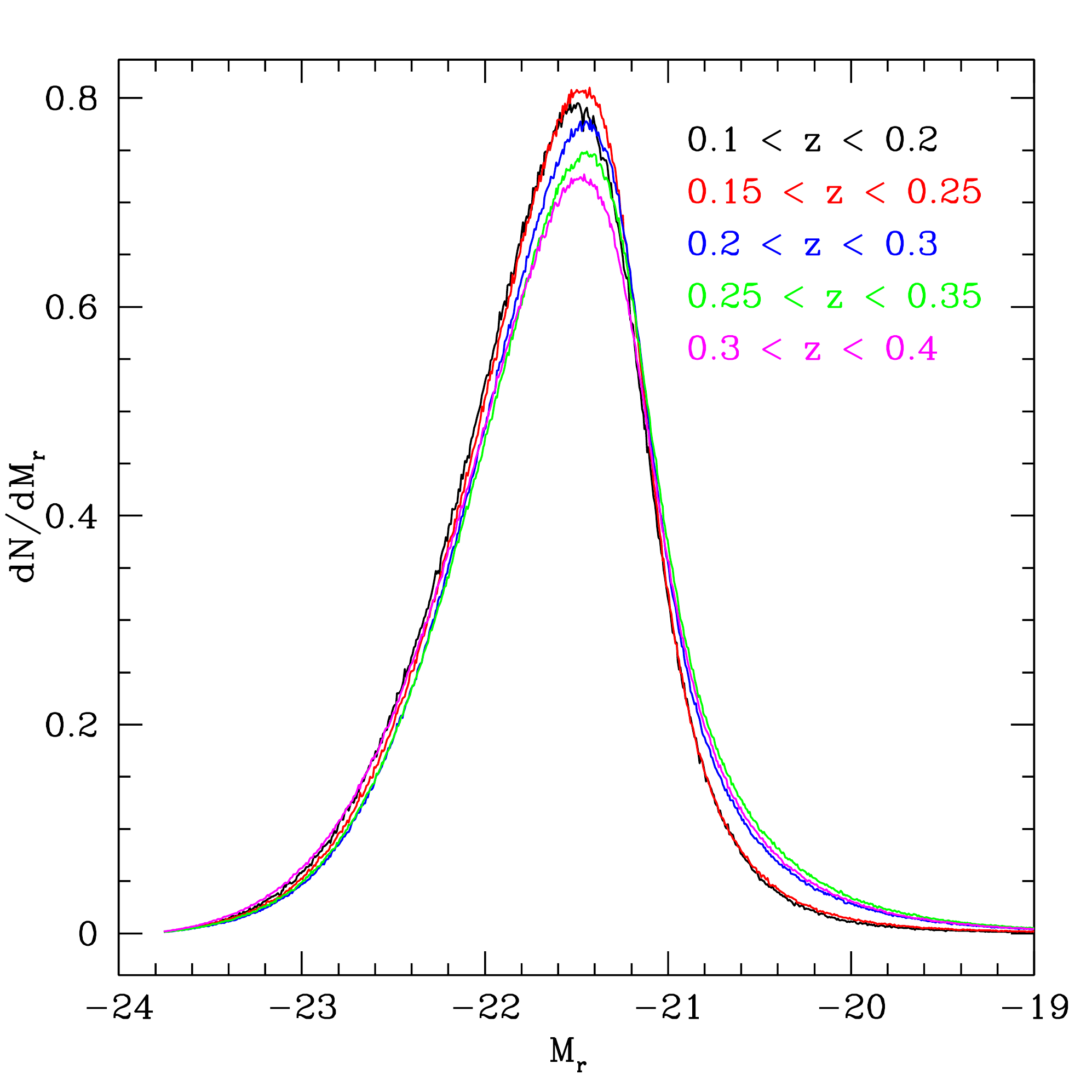}
\caption{The normalised (such that they integrate to 1) absolute $r$-band magnitude ($M_r$) distributions for each of the five photometric redshift slices $0.1 < z < 0.2 $ (black), $0.15 < z < 0.25 $ (red), $0.2 < z < 0.3 $ (blue), $0.25 < z < 0.35 $ (green), and $0.3 < z < 0.4$ (magenta).}
\label{fig:magdist}
\end{figure}

We can estimate the true absolute magnitude distributions in a manner that is similar to our estimation of $dN/dz$.  We follow the procedure outlined at the beginning of this section, but bin in $M_r$ instead of redshift.  It is important to consider the true distribution of $M_r$ in each of our samples because one may worry that applying the same cut on absolute magnitude to samples with differing photometric redshift errors could yield significantly different magnitude distributions.  Figure \ref{fig:magdist} displays the normalised (such that they integrate to 1) $M_r$ distributions for each of our samples ($0.1 < z < 0.2$, black; $0.15 < z < 0.25$, red; $0.2 < z < 0.3$, blue; $0.25 < z < 0.35$, green; and $0.3 < z < 0.4$, magenta).  The peak of each of the distributions occurs near $M_r = -21.5$ and there are only slight differences between the samples.  The higher redshift samples display slightly broader distributions (due to the fact that they have larger photometric redshift errors) and the broadening occurs most prominently at the faint end of the distributions.  This suggests, that, if the differences in these distributions cause any effect at all, it will be to slightly decrease the bias of the higher redshift samples. 

\subsection{Estimating True Completeness}
As in Ross \& Brunner (2009; hereafter R09), we use the redshift distribution to calculate the weighted number density, $n_g$, of observed galaxies using
\begin{equation}
n_g =\int {\rm d}z  \frac{H(z)}{4\pi f_{obs} x^2(z) c}\frac{{\rm d}N}{{\rm d}z}\times \left(\frac{{\rm d}N}{{\rm d}z}\right)^2 ~  /  ~ \int {\rm d}z\left(\frac{{\rm d}N}{{\rm d}z}\right)^2,
\label{eq:ngal}
\end{equation}
where $x(z)$ is the comoving distance to redshift $z$, $f_{obs}$ is the fraction of observed sky, $c$ is the speed of light, $H(z)$ is the rate of expansion, and $dN/dz$ is the un-normalised redshift distribution.  Equation (\ref{eq:ngal}) determines the number density of galaxies that contributes to a clustering measurement, and is therefore the best estimate of the observed number density of galaxies in a photometric redshift bin.

For our models, we need to know the true number density of galaxies for a given sample, and this is necessarily larger than the observed number density obtained from Equation (\ref{eq:ngal}).  If we assume our observed galaxies are a random sampling of this complete sample, the incompleteness should not affect our ability to model our measurements as long as we have a good estimate of the true number density.  We therefore assume the true number density of galaxies, $n_{gT}$, does not evolve with redshift and estimate $n_{gT}$ by taking the total number of galaxies in the $0.1 < z < 0.2$ sample and dividing it by the co-moving volume given by this redshift range.  This yields $n_{gT} = 0.0049 h^3$Mpc$^{-3}$.  We can thus compare the number density given by Equation (\ref{eq:ngal}) to 0.0049 $h^3$Mpc$^{-3}$ in order to estimate the completeness of each sample.  This will not be a concern for the modelling (which will always use $n_{gT} = 0.0049 h^3$Mpc$^{-3}$), but it will help us determine the quality of each of our data samples.

All of our samples will suffer incompleteness, in terms of the fraction of galaxies that contribute to the clustering signal, due to galaxies whose photometric redshift estimates have scattered them out of a particular redshift bin.  This effect is made clear by considering the redshift distributions of Figure \ref{fig:nzDR7}.  Each individual distribution clearly displays a significant portion of its curve lying outside bounds of its hard cut on photometric redshift (for example, the $dN/dz$ of the $0.3 < z < 0.4$ slice has significant amplitudes at $z < 0.3$ and $z > 0.4$).  Naturally, this effect grows larger as the mean redshift error of galaxies increases.

We display the number, the weighted number density ($n_g$), the median redshift ($\bar{z}$), the mean redshift error of galaxies ($\bar{\Delta z}$), the median $r$-band absolute magnitude ($M_r$), the completion (i.e. $n_g$/0.0049), and the weighted fraction of galaxies with type value $t > 0.1$ ($f_{late}$) for each redshift slice in Table 1.  The $\bar{M_r}$ are calculated from the magnitude distributions displayed in Figure \ref{fig:magdist}.  As the figure suggests, the values of $\bar{M_r}$ are extremely similar each other across all samples --- to the nearest tenth of a magnitude they are all equal to -21.6.  This suggests that we are comparing galaxies of the same luminosity in every sample.

The sharp decrease in completion for galaxies with $z > 0.2$ is coupled with a sharp increase in the mean redshift error.  For our sample with the highest redshifts ($0.3 < z < 0.4$), there is a significant decrease in the completion (from 0.48 to 0.42), but the increase in mean redshift error is not as significant (0.051 to 0.054).  This suggests that the data may be suffering from incompletion in the parent $r_d < 21$ sample, due to, e.g., low surface brightness objects (which \citealt{Blanton05sb} suggests may be an issue) and also the effects of Malmquist bias, since our $z < 0.4$ limit is imposed partly due to our $r_d < 21$ limit.    

We calculate the weighted late-type fraction by finding $n_{g,late}$ via Equation (\ref{eq:ngal}) (using $dN/dz_{late}$) and dividing the result by $n_g$.  We split the data at $t = 0.1$, since this split yields similar late-type fractions as the split used by R09.  The value of $f_{late}$ is important, as it affects both the overall bias and the shape of $\omega(\theta)$.  For all of the data with $z < 0.35$, the $f_{late}$ are consistent to within 10\% (and there is no overall trend).  There is, however, a large decrease in $f_{late}$ in the highest redshift slice.  This suggests that the large decrease in completeness in this slice may be tied to a deficit of late-type galaxies in our sample.  We should therefore be careful when comparing measurements from the $0.3 < z < 0.4$ slice to the lower-redshift ones.  Conversely, the agreement between the slices with $z < 0.35$ encourages comparisons between these four samples.

\begin{table*}
\begin{minipage}{7in}
\centering
\caption{Characteristics of the five SDSS DR7 samples with $M_r < -21.2$ used to calculate angular auto-correlation functions; $n_g$ is the weighted observed number density, $\bar{z}$ is the median redshift, $\bar{\Delta z}$ is the mean redshift error, $f_c$ is the completion ($n_g$/0.0049), and $f_{late}$ is the weighted fraction of late-type galaxies.  } 
\begin{tabular}{lccccccc}
\hline
\hline
Redshift range  & Number of galaxies &   $n_g$ (h$^3$/Mpc$^3$) & $\bar{z}$ & $\bar{\Delta z}$ & $\bar{M_r}$ & $f_c$ & $f_{late}$\\
\hline
$0.1 < z < 0.2$ &  483,655 & 0.00377 & 0.16 & 0.022 & -21.63 & 0.77  & 0.38\\
$0.15 < z < 0.25$ &  771,681 & 0.00369 & 0.2& 0.027 & -21.60 & 0.75 & 0.35\\
$0.2 < z < 0.3$ &  1,027,754 & 0.00272 & 0.25 & 0.041 & -21.57 & 0.56 & 0.35\\
$0.25 < z < 0.35$ &  1,406,302 & 0.00236 & 0.3 & 0.051 & -21.56 & 0.48 & 0.37\\
$0.3 < z < 0.4$ &  1,612,078 & 0.00204 & 0.34 & 0.054 & -21.59 &  0.42 & 0.29\\
\hline
\label{tab:DR7z4}
\end{tabular}
\end{minipage}
\end{table*}

\section{Measurement and Analysis Tools}
\label{sec:tools}
\subsection{Measurement Techniques}
\label{sec:meas}
We calculate the angular auto-correlation function, $\omega(\theta)$, of galaxies using the \cite{LZ} estimator:
\begin{equation}
\omega(\theta) = \frac{DD(\theta) - 2DR(\theta) + RR(\theta)}{RR(\theta)},
\end{equation}
where $DD$ is the number of galaxy pairs, $DR$ the number of galaxy-random pairs, and $RR$ the number of random pairs, all separated by an angular distance $\theta \pm \Delta\theta$.  We will also measure angular cross-correlation functions, $\omega_x(\theta)$, for which the \cite{LZ} estimator is
\begin{equation}
\omega_x(\theta) = \frac{D_1D_2(\theta) - D_1R(\theta)  - D_2R(\theta)+ RR(\theta)}{RR(\theta)},
\end{equation}
where $D_1$ and $D_2$ represent separate data samples.  We can always employ the same random catalog (which includes 10 million points) since all of our galaxies have the same angular selection.

We use a jackknife method (e.g., \citealt{Scr02}), with inverse-variance weighting to estimate our errors and covariance matrix (e.g., \citealt{Mye07}).  The method is nearly identical to the method described in detail in R07.  The jackknife method works by creating many subsamples of the entire data set, each with a small part of the total area removed.  R07 found that 20 jack-knife subsamplings are sufficient to create a stable covariance matrix, and we find similar results for DR7.  These 20 subsamples are created by extracting a contiguous grouping of 1/20th of the unmasked pixels in 20 separate areas.  Our covariance matrix, $C_{\rm jack}$, is thus given by
\begin{equation}
\begin{array}{ll}
C_{i,j, {\rm jack} }=  C_{\rm jack}(\theta_i,\theta_j) & ~\\
= \frac{19}{20} \sum_{k=1}^{20}[\omega_{full}(\theta_i) - \omega_{k}(\theta_i)][\omega_{full}(\theta_j) - \omega_{k}(\theta_j)], & ~
\end{array}
\label{eq:JK}
\end{equation} 
\noindent where $\omega_{k}(\theta)$ is the value for the correlation measurement omitting the $k$th subsample of data and $i$ and $j$ refer to the $i^{th}$ and $j^{th}$ angular bin. The jackknife errors are simply the square-root of diagonal elements of the covariance matrix.  Such a technique should account for statistical errors due to variations in both the angular plane and the radial direction, as each jackknife represents a different realisation of the radial selection. 

\cite{Norberg08} have shown that using a jack-knife method to estimate covariance matrices does not yield perfect results.  For projected correlation function measurements (the case they study most similar to ours) the jack-knife method does well at large scales, but over-predicts the variance at small scales.  For the covariance, again the jack-knife method is shown to be imperfect.  We do not feel this is a major issue for the interpretation of our measurements as the conclusions we draw will not depend heavily on the exact nature of the covariance matrices.  We explore this further in section \ref{sec:robust}.

\subsection{Transformation to Angular Correlation Function}
Our theoretical modeling will produce galaxy-galaxy power spectra, $P(k,r)$.  Thus, we must Fourier transform the model power spectra to a real-space correlation function, $\xi(r)$,
\begin{equation}
\xi(r) = \frac{1}{2\pi^2}\int_{0}^{\infty}{\rm d}k~ P(k,r)k^2 \frac{{\rm sin}~ kr}{kr},
\end{equation}
where $r$ is the real-space distance and our model $P(k,r)$ will depend on it due to halo-exclusion (see section \ref{sec:HM}).  We use Limber's equation (\citealt{lim}) to project the real-space model to angular space (assuming a flat Universe):
\begin{equation}
\begin{array}{ll}
\omega(\theta) = &~\\ 
 2/c \int_{0}^{\infty}{\rm d}z ~H(z)(dN/dz)^2 \int_{0}^{\infty}{\rm d}u~\xi(r= \sqrt{u^2+x^2(\bar{z})\theta^2}), & ~
\end{array}
\label{eq:Lim}
\end{equation}
where $dN/dz$ is the normalised redshift distribution, and $x(\bar{z})$ is the comoving distance to the median redshift $\bar{z}$.  The factor $1/c \int_{0}^{\infty}{\rm d}z ~H(z)(dN/dz)^2$ essentially tells one how much the radial extent of the galaxy distribution dilutes the real-space clustering signal.  This will therefore change for each redshift sample, and it is thus an important factor when comparing results between different redshift slices.  We therefore define
\begin{equation}
W = 1/c \int_{0}^{\infty}{\rm d}z ~H(z)(dN/dz)^2,
\label{eq:wnz}
\end{equation}
and we will use this factor $W$ in order to enable direct comparison of our measurements to each other.

Recent studies (e.g. \citealt{Pad07, Bald, Nock}) have shown that redshift distortions can significantly affect projected correlation function measurements.  The importance of the redshift distortion effect grows with the effective scale, and at the scales we probe ($r_{eq}< 15h^{-1}$Mpc), it would increase our models by at most $\sim5\%$, given that the sizes of our radial windows are all greater than 250 $h^{-1}$Mpc.  This suggests that including the effects of redshift distortions would not alter our models significantly enough to alter any of our conclusions. 

\subsection{Passive Evolution}
In order to compare our measurements in different redshift shells, we must take into account the evolution of the clustering of the dark matter.  The overall growth of structure in the Universe implies that the bias, $b(z)$, of a passively evolving set of galaxies (i.e., no mergers) will tend towards unity.  Specifically, this is expressed as
\begin{equation}
b(z_1) = 1 +(b(z_2) - 1)D(z_2)/D(z_1),
\label{eq:b0}
\end{equation}
(see, e.g., Fry 1996, Tegmark \& Peebles 1998) where $D(z)$ is the linear growth factor.  Thus, assuming no evolution in the physical properties of individual galaxies, a passively evolving set of galaxies with $b = 1.40$ at $z = 0.34$ should have a bias of 1.335 today ($z=0$).  Given that $\frac{\xi(z_1)}{\xi(z_2)} = \left(\frac{b(z_1)D(z_1)}{b(z_2)D(z_2)}\right)^2$, one can express the ratio of the real-space clustering amplitude between two redshifts as
\begin{equation}
f_{\xi}(z_1,z_2) = \left(\frac{b(z_1)}{b(z_1)-1+D(z_2)/D(z_1)}\right)^2.
\label{eq:fxi}
\end{equation}

Given Equations (\ref{eq:wnz}) and (\ref{eq:fxi}), we can account for the expected changes in the angular clustering due to both the changes in the widths of the redshift distributions and the median redshifts.  Given the median redshift, $\bar{z}$, one can determine $\omega(r_{eq})$ by finding the equivalent physical scale of a given angular separation via
 \begin{equation}
 r_{eq} = 2x(\bar{z}){\rm tan}(\theta/2),
 \end{equation}
  (where again $x(\bar{z})$ is the comoving distance to median redshift $\bar{z}$).  Thus, the expected difference between $\omega(r_{eq})$ measured at $\bar{z}_1$ and $\bar{z}_2$ can be expressed as
\begin{equation}
f_{\omega}(\bar{z}_1,\bar{z}_2) = f_{\xi}W_1/W_2,
\label{eq:omeganorm}
\end{equation}
where $W_1$ is determined via Equation (\ref{eq:wnz}) for redshift distribution with median redshift $\bar{z}_1$.    Any difference between $f_{\omega}(\bar{z}_1,\bar{z}_2)$ and $\frac{\omega(r_{eq},\bar{z}_1)}{\omega(r_{eq},\bar{z}_2)}$ would thus be due evolutionary effects, such as the merging or dimming of galaxies.

 \subsection{Halo Modeling}
 \label{sec:HM}
We use the halo model to produce model galaxy auto-correlation functions using techniques similar to those outlined in R09.  We assume that the overall galaxy power spectrum can be modeled as having a contribution due to galaxy pairs within dark matter halos (the `1-halo' term) and a contribution due to galaxies pairs in separate halos (the `2-halo' term).  The number of galaxies expected to occupy a halo is modeled as a function of mass, and the 1 and 2-halo terms can be self-consistently determined given this HOD.  

As in previous studies (see, e.g., \citealt{zheng05,Blake}; R09), we assume separate mean occupations for central galaxies, $N_{c}(M)$ and for satellite galaxies, $N_{s}(M)$.  Thus,
\begin{equation}
N(M) = N_c(M)\times(1 + N_s(M)),
\label{eq:NHOD}
\end{equation} 
where we are assuming that only halos with central galaxies can have satellite galaxies.  This allows for two 1-halo components --- one for central-satellite pairs, $P_{cs}(k)$, and the other for satellite-satellite, $P_{ss}(k)$, pairs.  They are given by (see, e.g., Appendix B of \citealt{skibbasheth09})
\begin{equation}
P_{cs}(k) = \int^{\infty}_{M_{vir}(r)} dM n(M) N_c(M)  \frac{2  N_s(M ) u(k|M)}{n^2_{gT}},
\label{eq:cs}
\end{equation}

\begin{equation}
P_{ss}(k) =\int^{\infty}_{0} dM n(M) N_c(M) \frac{ \left(N_s(M)u(k|M)\right)^2}{n^2_{gT}},
\label{eq:ss}
\end{equation}
where the factor $n(m)$ is the halo number density, for which we use the \cite{Jenk01} model (and is implicitly dependent on redshift), and $u(k|M)$ is the Fourier transform of the \cite{NFW}  (NFW) dark matter profile.  We have implicitly assumed that the satellite galaxies are poisson distributed (as is found to be a good approximation by, e.g., \citealt{krav04,zheng05}) allowing the use of $N_s(M)^2$ in place of $\langle N_s(M)(N_s(M) - 1)\rangle$.

The 2-halo term is given by
\begin{equation}
\begin{array}{ll}
P_{2h}(k,r) & =  P_{matter}(k) \\
~ & \times \left [ \int_0^{M_{lim}(r)} dM n(M) b(M,r) \frac{N(M)}{n_g^{\prime}}u(k|M) \right]^2, 
\end{array}
\end{equation}
where $P_{matter}$ is the matter power-spectrum determined via the fitting formulae of \cite{Smith} and $b(M,r)$ is the scale dependent bias of halos.  This bias can be expressed \citep{Tinker} as 
\begin{equation}
b^2(M,r) = B^2(M) \frac{[1+1.17\xi_m(r)]^{1.49}}{[1+0.69\xi_m(r)]^{2.09}},
\end{equation}
 where $B(M)$ is the bias of halos, which we calculate using the model of \cite{Sheth01} with the parameterisation determined by \cite{Tinker} (and is implicitly dependent on redshift) and $\xi_m(r)$ is the non-linear real-space matter 2-point correlation function, determined by Fourier transforming the matter power spectrum.  The parameter $M_{lim}(r)$ is the mass limit due to halo-exclusion, which we determine using the methods described by \cite{Tinker} and \cite{Blake}.  
 
One can calculate the average number density of galaxies for a given HOD, $n_{gH}$, via 
\begin{equation}
n_{gH} = \int_0^{\infty}{\rm d}Mn(M)N(M),
\label{eq:ng}
\end{equation}
and the {\it restricted} number density of galaxies, $n^{\prime}_{g}$, via
\begin{equation}
n^{\prime}_{g} = \int_0^{M_{lim}(r)}{\rm d}Mn(M)N(M).
\end{equation}
The full model galaxy-galaxy power spectrum is thus given by $P(k,r) = P_{cs}(k) + P_{ss}(k) + P_{2h}(k,r)$.  
 
 For the central galaxy HOD, we use the same parameterisation as in R09, i.e.,
 \begin{equation}
N_c(M) = 0.5 \left[ 1 + {\rm erf}\left(\frac{{\rm log_{10}} (M/M_{cut})}{\sigma_{cut}}\right)\right] . 
\label{eq:HODc}
\end{equation}
For the satellite galaxy HOD, we use 
\begin{equation}
N_s(M) =  \left(\frac{M-M_{cut}}{M_1}\right)^{\alpha} . 
\label{eq:HODs}
\end{equation}
This is similar to the parameterisation used by \cite{ZZ07}, $N_s(M) = \left(\frac{M-M_o}{M_1}\right)^{\alpha}$, but we use $M_{cut}$ instead of adding a new $M_o$ parameter.  This is motivated by the fact that, in \cite{ZZ07}, $M_o$ is loosely constrained, but consistent with $M_{cut}$ for each of their SDSS samples.  Our parameterisation thus allows for a physically motivated form for the satellite HOD, without adding an extra parameter into the model.  The total mean occupation of halos at a given mass is therefore determined by entering Equations (\ref{eq:HODc}) and (\ref{eq:HODs}) into Equation (\ref{eq:NHOD}).

The HOD model has four free parameters, but one can be removed by requiring that $n_{gH}$ calculated via Equation (\ref{eq:ng}) matches the observed number density of galaxies, $n_{gT}$ (which we take to be 0.0049 $h^3$Mpc$^{-3}$ for each sample; see \S \ref{sec:data}).  Thus, given $\alpha$, $M_1$, and $\sigma_{cut}$, we find the $M_{cut}$ that yields a match between $n_{gT}$ and $n_{gH}$.  For comparison purposes, we will wish to know the linear bias, $b_1$, and the satellite fraction, $f_{sat}$, intrinsic to a given HOD.  These can be expressed as
\begin{equation}
b_1 = \frac{1}{n_g} \int {\rm d}M B(M) n(M)N(M) 
\label{eq:b1}
\end{equation}
and
\begin{equation}
f_{sat} = \frac{1}{n_g} \int {\rm d}M n(M)N_c(M)N_s(M) .
\label{eq:fs}
\end{equation}

\section{Clustering Measurements and Best-fit HODs}
\label{sec:res}
\begin{figure}
\includegraphics[width=84mm]{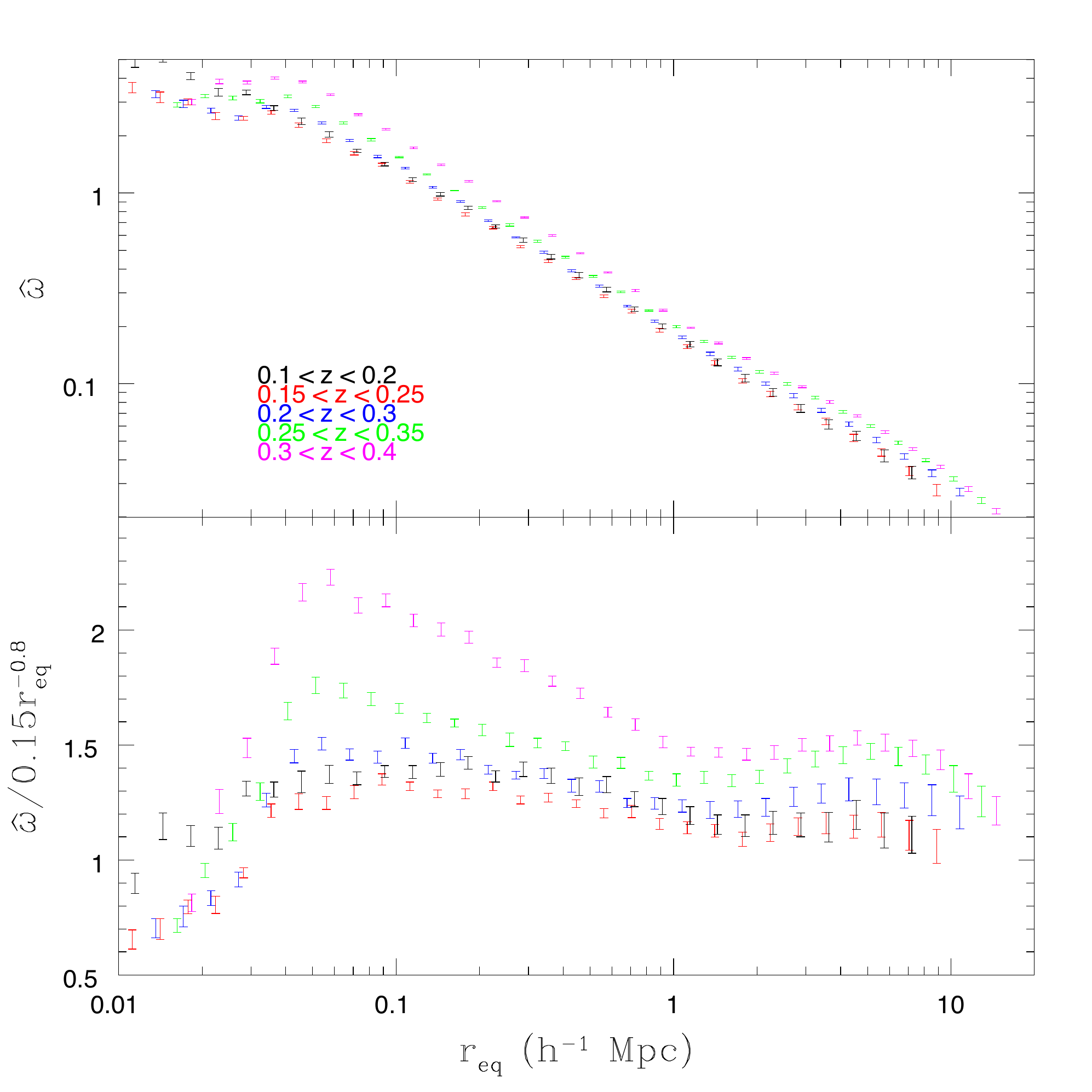}
\caption{Top panel: the measured (error bars) angular auto-correlation functions for five photometric redshift slices $0.1 < z < 0.2$ (black), $0.15 < z < 0.25$ (red), $0.2 < z < 0.2$ (blue), $0.25 < z < 0.35$ (green), $0.3 < z < 0.4$ (magenta).  The amplitudes have been normalized to account for changes in the redshift distributions and passive evolution and are plotted against the equivalent physical scale.  Bottom panel:  the same information as the top panel, only the $\hat{\omega}$ values have been divided by the power-law $0.15r_{eq}^{-0.8}$.}
\label{fig:w2norm}
\end{figure}

\begin{figure}
\includegraphics[width=84mm]{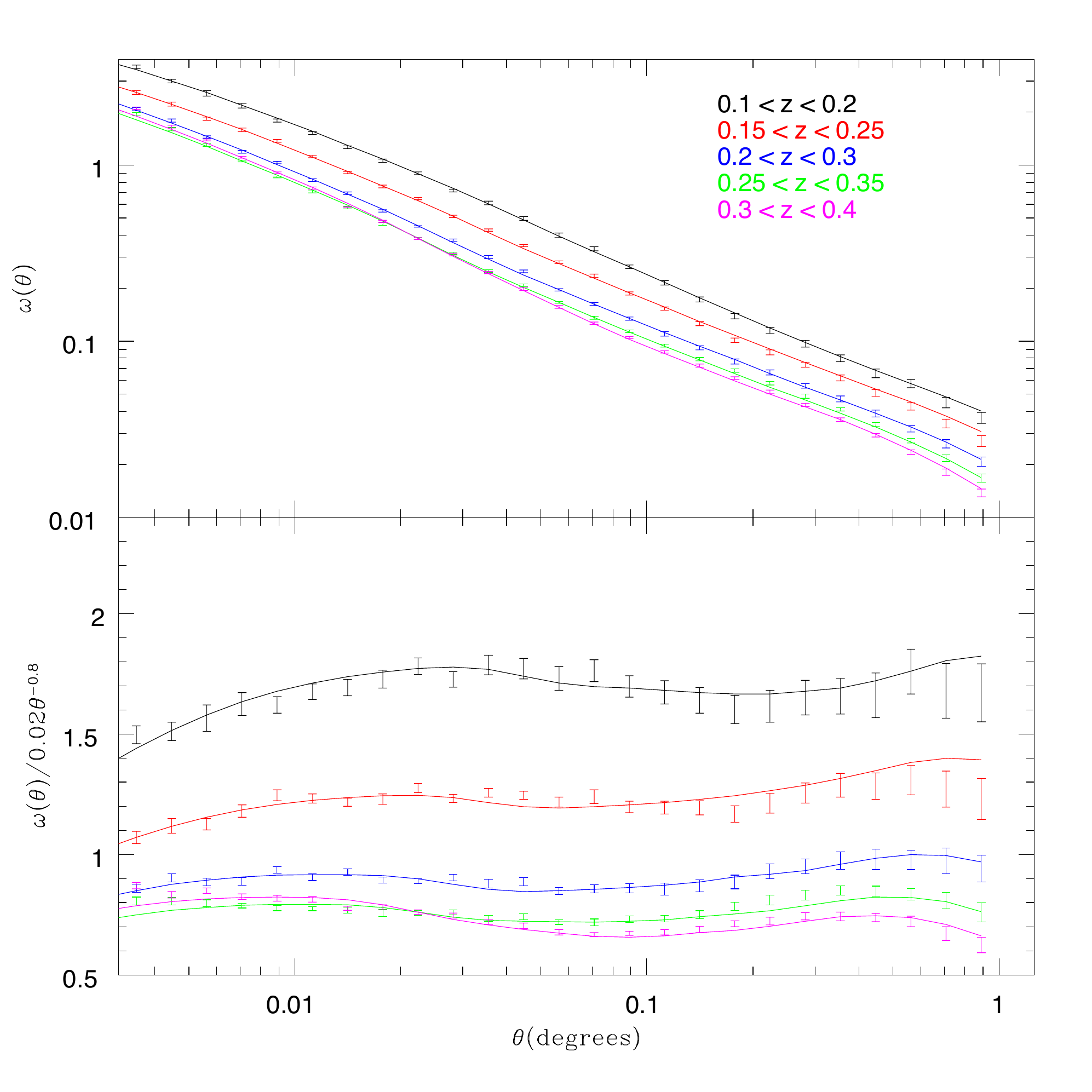}
\caption{Top panel: the measured (error bars) and model (solid lines) angular auto-correlation functions for five photometric redshift slices $0.1 < z < 0.2$ (black), $0.15 < z < 0.25$ (red), $0.2 < z < 0.2$ (blue), $0.25 < z < 0.35$ (green), $0.3 < z < 0.4$ (magenta).  Bottom panel:  same information as the top, except that the amplitudes of $\omega(\theta)$ are divided by the power-law 0.02$\theta^{-0.8}$.}
\label{fig:res}
\end{figure}

Our measured angular auto-correlation functions, normalised using Equation (\ref{eq:omeganorm}), are plotted (error-bars) against the equivalent physical scale in the top panel of Figure \ref{fig:w2norm} for each of our redshift slices ($0.1 < z < 0.2$ black,  $0.15 < z < 0.25$ red, $0.2 < z < 0.3$ blue, $0.25 < z < 0.35$ green, $0.3 < z < 0.4$ magenta).  We use the bias of best-fit HOD model for the $0.2 < z < 0.3$ sample (1.246; see Table \ref{tab:res}) as the $b(z_1)$ that enters Equation (\ref{eq:fxi}) (and then Equation (\ref{eq:omeganorm})), since this model is most consistent with the measurements at large angular scales (see Figure \ref{fig:res}).   All of the measurements display a turnover at $\sim50h^{-1}$Kpc, suggesting a minimum physical scale below which we can not observe a pair of $M_r < -21.2$ galaxies.  Looking specifically at $0.3 < z < 0.4$ sample, the measured amplitudes at small scales are significantly higher than any of the other samples, suggesting that the low completion and late-type fractions of this sample have indeed influenced the resulting measurements.

The amplitudes of the measurements in Figure \ref{fig:w2norm} clearly grow larger as the redshift increases.  This is made most clear by looking at the bottom panel of Figure \ref{fig:w2norm}, where we plot the same information as the top panel, but divide the amplitudes by the power-law $0.15r_{eq}^{-0.8}$.  If our sample of galaxies evolved passively, all of the measurements in Figure \ref{fig:w2norm} would be consistent with each other (in both panels).  This suggests that there is significant evolution in the properties of galaxies with $M_r < -21.2$, and that this evolution causes lower redshift galaxies to exist in significantly less biased environments than their high-redshift counterparts.      

The top panel of Figure \ref{fig:res} plots (same colour scheme as Figure \ref{fig:w2norm}) displays our measured auto-correlation functions (without any normalisation) along with the best-fit model for each measurement, while the bottom panel displays the same information divided by the power-law $0.02\theta^{-0.8}$.  By eye, the fits appear good, and the greatest disagreement appears to be at large angular scales for each measurement.  This is due, in part, to the fact that the error-bars are largest at these scales.  For the higher redshift samples, the largest disagreement is around scales $\sim$4$h^{-1}$Mpc.  This is large enough that the result is predominantly dependent on the 2-halo term.  This hints that our assumed cosmology may be off, but such considerations are beyond the scope of this paper. 

\subsection{Best-fit HODs}
The best-fit HOD parameters for each redshift slice are presented in Table \ref{tab:res}.  For each measurement, we fit the model between 0.004$^{\rm o}$ and 1.0$^{\rm o}$.  There are 24 measurements over this range and thus 21 degrees of freedom for each fit.  For the two highest redshift ranges, the value of $\chi^2/\nu$ is greater than 1 (but never greater than 2).  We do not think that this suggests a problem, because the covariance only takes the statistical error of the correlation function measurements into account.  These two samples have the most data, their statistical errors are thus the smallest.  There are many other sources of potential error, such as the assumed cosmology and the fitting formulae utilised by the models.  

Our results are slightly different to those of R09, who used galaxies with $0 < z < 0.3$ photometrically selected from SDSS DR5 with $M_r -5{\rm log}(h) < -20.5$.  Their best-fit $M_{cut}$ and $\sigma_{cut}$ are similar to ours, but their best-fit $\alpha$ and $M_1$ are significantly higher (our are $\alpha \sim 1.1$ and log$_{10}(M_1 h/M_{\odot}) \sim 13.3$ while R09 found $\alpha = 1.27$ and log$_{10}(M_1 h/M_{\odot}) = 13.49$).  The main reason for this is that they used a different treatment for $N_{sat}$, and this treatment naturally results in best-fits with both larger $\alpha$ and $M_1$ values.  Another consequence of this change in the modeling is that the resulting satellite fractions are higher than those of R09.  

Compared to \cite{ZZ07}, the best-fit values of our HOD parameters mainly fall between their results for SDSS galaxies with $M_r -5{\rm log}(h) < -20$ and $M_r  -5{\rm log}(h) < -20.5$.  This makes sense given that our cutoff is at $M_r  -5{\rm log}(h) < -20.43$ and it is made effectively less luminous by the fact that photometric redshift errors allow some lower luminosity galaxies into our samples.  Comparing our best-fit values of $M_1$ and $M_{cut}$, we find that $M_1/M_{cut}$ varies between 12.2 and 16.6, but is less than 13.8 only for the $0.3 < z < 0.4$ sample.  Theoretical predictions by \cite{zheng05} found the ratio to be $\sim 14$ and $\sim 18$ using two separate galaxy formation models, and our results fall in between these.  

As expected from Figure \ref{fig:w2norm}, the bias values of the best-fit HOD models increase with redshift.  The driving factor behind this change is the value of $\sigma_{cut}$, as when its best-fit value drops, the bias increases.  This is due to the fact that a smaller value of $\sigma_{cut}$ results in a sharper cutoff in the HOD profile, fewer galaxies in low-mass halos, and thus a larger value for the overall bias.  The change in the bias is significantly larger than the change expected from passive evolution.  Given a $b_1$ of 1.2 at $z = 0.16$, the bias of a passively evolving sample is 1.22 at $z = 0.34$ --- $\sim 10\%$ smaller than the bias value of our best-fit HOD.

\begin{table*}
\centering
\begin{minipage}{7in}
\caption{The best-fit values of the HOD parameters (see Equations (\ref{eq:HODc}) and (\ref{eq:HODs}) )and $1\sigma$ errors and the associated $\chi^2$ values for the five redshift slices studied, fit between 0.004$^{\rm o}$ and 1$^{\rm o}$.  All masses are in units $M_\odot h^{-1}$.  The parameters $b_1$ and $f_{sat}$ are the linear bias and the satellite fraction, given the best-fit HOD parameters and calculated using Equations (\ref{eq:b1}) and  (\ref{eq:fs}), respectively.} 
\begin{tabular}{lccccccc}
\hline
\hline
Redshift Range & $\alpha$ & ${\rm log_{10}}\left(M_{cut}\right)$ &  ${\rm log_{10}}\left(M_{1}\right)$  & $\sigma_{cut}$ & $\chi^2/{\rm dof}$ &   $b$ &   $f_{sat}$\\
\hline
$0.1 < z < 0.2$ & 1.093$^{+0.012}_{-0.015}$ & 12.162 & 13.303$\pm$0.008 & 0.49$^{-0.09}_{+0.05}$ &  10.8/21  & 1.187 & 0.203  \\
$0.15 < z < 0.25$ & 1.103$\pm$0.008 & 12.154 & 13.319$\pm$0.005 & 0.49$^{+0.06}_{-0.04}$ &  17.4/21  & 1.19 & 0.190  \\
$0.2 < z < 0.3$ & 1.100$^{+0.008}_{-0.01}$ & 12.089 & 13.300$\pm$0.005 & 0.33$\pm$0.04 &  19.0/21  & 1.231 & 0.199  \\
$0.25 < z < 0.35$ & 1.096$\pm0.004$ & 12.051 & 13.254$\pm$0.002 & 0.23$\pm$0.02 &  31.9/21  & 1.286 & 0.212  \\
$0.3 < z < 0.4$ & 1.075$\pm$0.005 & 12.064 & 13.177$\pm$0.003 & 0.2$\pm$0.03 &  33.6/21  & 1.347 & 0.245 \\
\hline
\label{tab:res}
\end{tabular}
\end{minipage}
\end{table*}

\begin{figure}
\includegraphics[width=84mm]{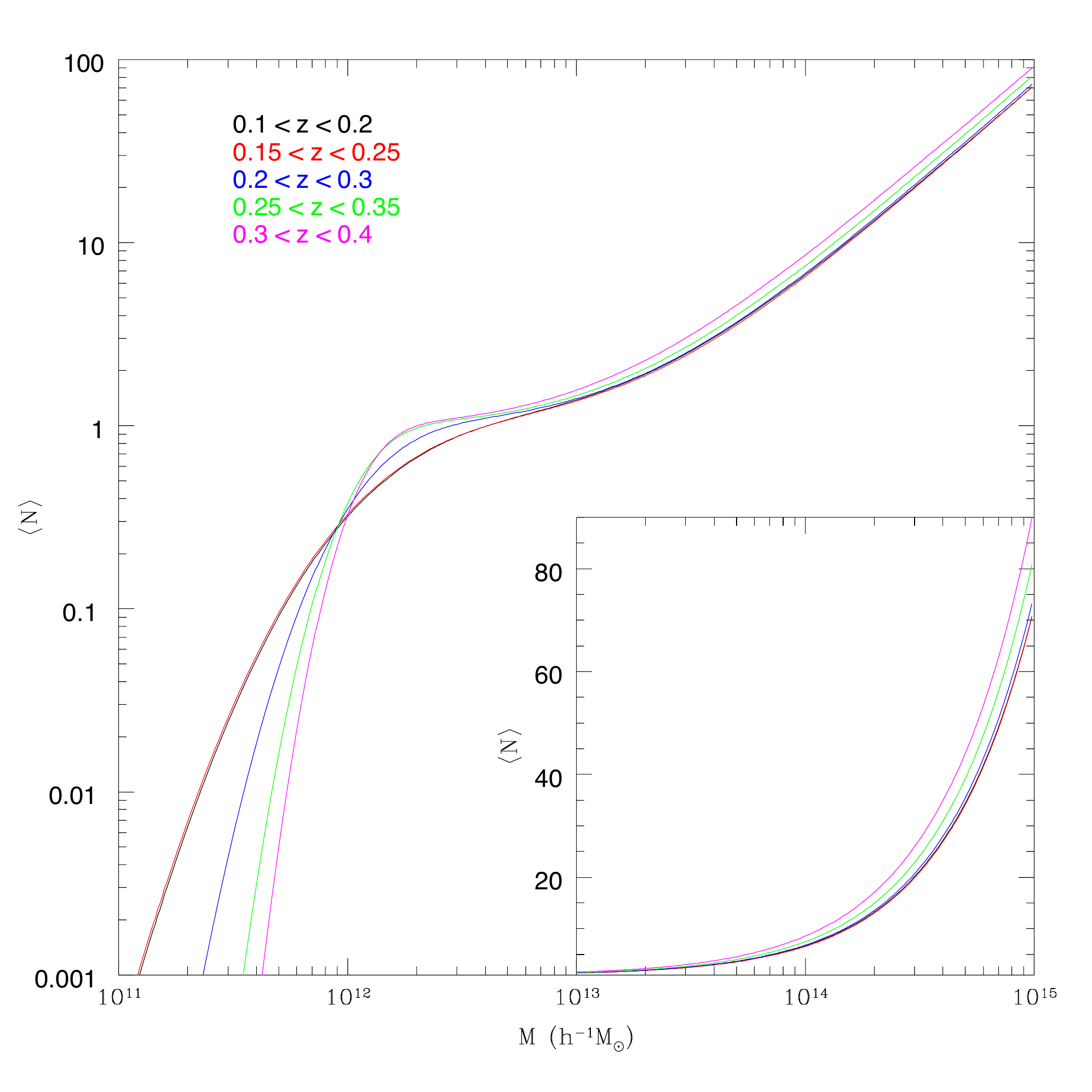}
\caption{The best-fit HOD models for the five photometric redshift slices $0.1 < z < 0.2$ (black), $0.15 < z < 0.25$ (red), $0.2 < z < 0.2$ (blue), $0.25 < z < 0.35$ (green), $0.3 < z < 0.4$ (magenta).  The inset plot displays the same data at high mass with $\langle N\rangle$ scaled linearly.}
\label{fig:HOD}
\end{figure}

Figure \ref{fig:HOD} plots the best-fit HOD for each measurement (same colour scheme as Figure \ref{fig:w2norm}).  The largest differences are at the low mass end, where best-fit value of $\sigma_{cut}$ affects the shape of the HOD by controlling the sharpness of the mass cutoff.  The best-fit value of $M_{cut}$ remains roughly constant (log$_{10}( M_{cut} ~h/M_{\odot})$ is always between 12.05 and 12.16) allowing the value of $\sigma_{cut}$ to be the dominant factor in the shapes of the best-fit HODs.  The trend we find in $\sigma_{cut}$ is not completely unexpected, as the change in distance modulus is larger across lower redshift bins and thus a wider range of luminosities may contribute to the mass cut-off scale.  One might be more comfortable with the results, however, if the change in $\sigma_{cut}$ were not so large.  We investigate this further in \S \ref{sec:sigcut}.  

At high mass, the HOD models look similar, though the high-redshift haloes host slightly more galaxies per halo.  This change is due to the best-fit value of $M_1$, which we find to decrease with redshift for the four samples with $z > 0.15$.  The decrease is rather dramatic between the $0.25 < z < 0.35$ and the $0.3 < z < 0.4$ samples (log$_{10}(M_1~ h/M_{\odot})$ decreases from 13.244 to 13.177), which causes the $0.3 < z < 0.4$ best-fit HOD to be significantly higher at the high mass end than any other sample.  This also has a consequence for the satellite fraction corresponding to the best-fit HOD.  For the four samples with $z < 0.35$, the satellite fraction stays within $ 6\%$ of 0.2 and there is no trend.  These satellite fractions that we find are similar to those found by \cite{ZZ07} for SDSS galaxies with $M_r-5{\rm log}(h) < -20.5$.  For the $0.3 < z < 0.4$ sample, the satellite fraction jumps to 0.245.  We discuss this further in \S \ref{sec:EL}.  Across all of our samples, the best-fit value of $\alpha$ has no trend, and it is consistent enough that it affects no noticeable difference in the shape of the best-fit HODs.  

\subsubsection{Fixed $\sigma_{cut}$}
\label{sec:sigcut}
\begin{figure}
\includegraphics[width=84mm]{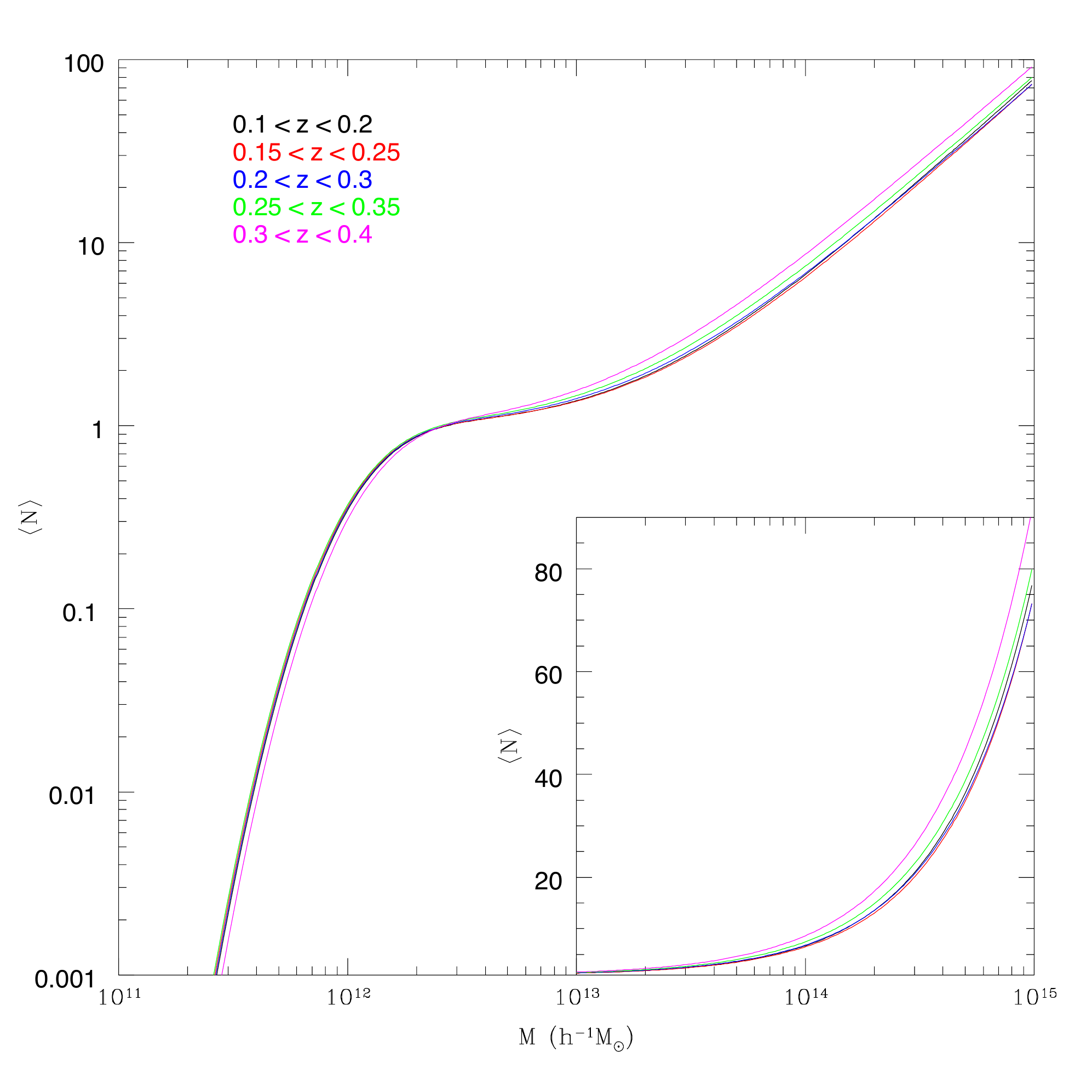}
\caption{Same as Figure \ref{fig:HOD}, only for $\sigma_{cut} = 0.3$.}
\label{fig:HODplotfixsig}
\end{figure}

\begin{table*}
\centering
\begin{minipage}{7in}
\caption{Same as Table \ref{tab:res}, only the parameter $\sigma_{cut}$ is fixed at 0.3.  All masses are in units $M_\odot h^{-1}$.} 
\begin{tabular}{lcccccc}
\hline
\hline
Redshift Range & $\alpha$ & ${\rm log_{10}}\left(M_{cut}\right)$ &  ${\rm log_{10}}\left(M_{1}\right)$   & $\chi^2/{\rm dof}$ &   $b_1$ &   $f_{sat}$\\
\hline
$0.1 < z < 0.2$ & 1.13 & 12.083 & 13.326  &  11.9/22  & 1.205 & 0.204  \\
$0.15 < z < 0.25$ & 1.126 & 12.075 & 13.339  &  23.1/22  & 1.206 & 0.190  \\
$0.2 < z < 0.3$ & 1.10 & 12.080 & 13.30  &  19.8/22 & 1.233 & 0.200  \\
$0.25 < z < 0.35$ & 1.093 & 12.070 & 13.254  &  34.9/22  & 1.286 & 0.209  \\
$0.3 < z < 0.4$ & 1.078 & 12.089 & 13.175  &  38.0/22  & 1.346 & 0.243 \\
\hline
\label{tab:sigfix}
\end{tabular}
\end{minipage}
\end{table*}

The differences in the best-fit HOD for each redshift range are driven in large part by the changes in $\sigma_{cut}$.  The best-fit values of $\sigma_{cut}$ generally decrease with redshift and thus cause a sharper mass cutoff at higher redshift.  To investigate whether our results are potentially biased by the changes in this parameter (which may or may not be physical) we therefore fix $\sigma_{cut} = 0.3$ and re-calculate the best-fit HODs for each redshift slice, with these best-fit parameters presented in Table \ref{tab:sigfix}.  The increase in the best-fit $\chi^2$ value is greater than 1$\sigma$ only for the $0.15 < z < 0.25$ and $0.3 < z < 0.4$ slices.  The increase in the $\chi^2$ values for the lower redshift samples is due to the fact that the lower $\sigma_{cut}$ value forces the overall bias to be larger and thus less consistent with our measurements at large angular scales.

Figure \ref{fig:HODplotfixsig}, displays the best-fit HODs for $\sigma_{cut} = 0.3$.  They look extremely similar to each other --- only the $0.3 < z < 0.4$ best-fit HOD is distinguishable from the other curves.  This suggests that the spread in best-fit HODs that we obtain when we leave $\sigma_{cut}$ as a free parameter is caused by the uncertainty inherent in our best-fit HODs, i.e., there is some degeneracy between $\sigma_{cut}$ and $M_{cut}$.  This also illustrates the fact that having the same HOD at two different redshifts actually implies quite different clustering.  The $0.1 < z < 0.2$ and $0.25 < z < 0.35$ best-fit HODs are nearly identical, yet the bias is 10$\%$ higher for the $0.25 < z < 0.35$ best-fit HOD.  We discuss the implications of this further in \S \ref{sec:phys}.   
 
\subsection{Splitting by Type}
\label{sec:EL}
\begin{figure*}
\begin{minipage}{7in}
\centering
\includegraphics[width=180mm]{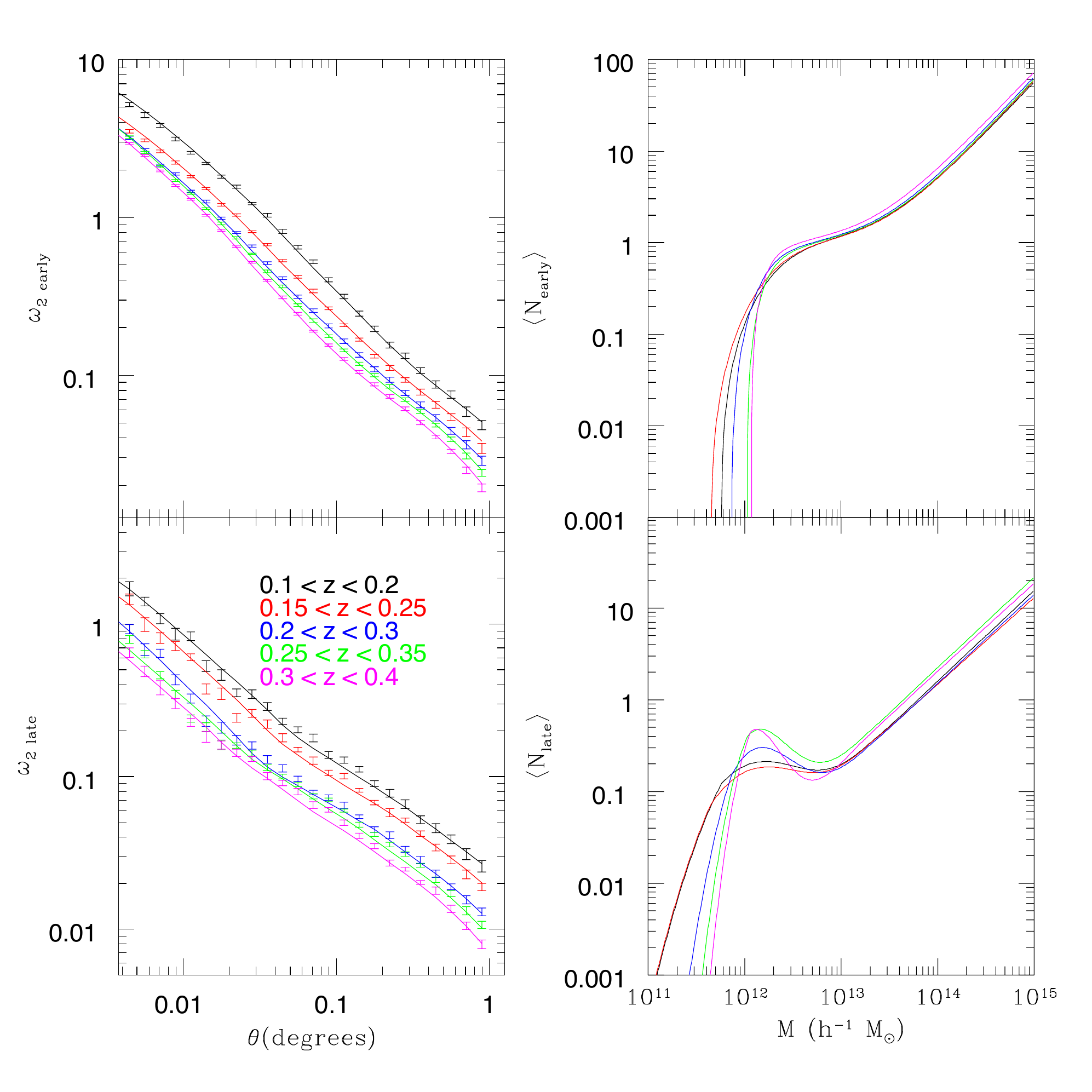}
\caption{Left panels: the measured (error bars) angular 2-point correlation functions for five photometric redshift slices $0.1 < z < 0.2$ (black), $0.15 < z < 0.25$ (red), $0.2 < z < 0.2$ (blue), $0.25 < z < 0.35$ (green), $0.3 < z < 0.4$ (magenta) for early-type (top) and late-type galaxies (bottom), compared against the best-fit model $\omega(\theta)$.  Right panels: the best-fit HOD models for the early- (top) and late-type galaxies (bottom; same colour scheme as left-hand panels)}
\label{fig:z4EL}
\end{minipage}
\end{figure*}

\begin{table*}
\centering
\begin{minipage}{10in}
\caption{The best-fit values of the HOD parameters and the associated $\chi^2$ values for the early- and late-type samples studied.} 
\begin{tabular}{lccccccc}
\hline
\hline
Sample  & $f_{c0}$ &  $f_{s0}$&  $\sigma_{cen}$ &   $\sigma_{sat}$&   $\chi^2/{\rm dof}$ &  $b_{1, late}$ &   $b_{1, early}$  \\
\hline
$0.1 < z < 0.2$ &  0.42 & 0.33$\pm$0.01 & 0.47$\pm$0.03  & 4.1$\pm$0.6 & 61/44 &  1.02 & 1.28 \\
$0.15 < z < 0.25$ &  0.36 & 0.34$\pm$0.01 & 0.50$\pm$0.04  & 2.7$\pm$0.5 & 79/44 &  1.02 & 1.28\\
$0.2 < z < 0.3$ &  0.57 & 0.31$\pm$0.01 & 0.39$\pm$0.03  & 3.2$\pm$0.4 & 60/44 &  1.06 & 1.32\\
$0.25 < z < 0.35$ &  0.88 & 0.42$\pm$0.01 & 0.34$\pm$0.03  & 3.8$\pm$0.4 & 109/44 &  1.13 & 1.39\\
$0.3 < z < 0.4$ &  0.93 & 0.34$\pm$0.01 & 0.22$^{+0.04}_{-0.03}$  & 3.6$\pm$0.3 & 70/44 &  1.16 & 1.43\\
\hline
\label{tab:HODel}
\end{tabular}
\end{minipage}
\end{table*}

As in R09, we can gain insight by splitting our sample by type value and measuring the respective auto-correlation functions.  This allows us to determine if the trends we observe are driven by a single galaxy type.  In this work, we refer to galaxies with $t > 0.1$ as late-type and those with $t < 0.1$ as early-type.  This split is motivated by the fact that, at low-redshift, it yields similar distributions as a $u-r = 2.2$ split in colour (as suggested by \citealt{Strat01}; the type-value split should perform better as a function of redshift than this simple colour cut).  We fit HOD models to the early- and late-type $\omega(\theta)$ measurements by assuming that the fraction of late-type galaxies that are central galaxies, $f_c$, exhibits a decrease with mass parameterised by
\begin{equation}
f_c(M) = f_{c0} ~ {\rm exp}\left[\frac{ -{\rm log_{10}}(M/M_{cut})}{\sigma_{cen}}\right] ,
\label{eq:cent}
\end{equation}
and the fraction of satellite galaxies that are late-type, $f_s$, exhibits a decrease with mass parameterised by
\begin{equation}
f_s(M) = f_{s0} ~ {\rm exp}\left[\frac{ -{\rm log_{10}}(M/M_{0})}{\sigma_{sat}}\right] .
\label{eq:sat}
\end{equation}
As in R09, we find $f_{c0}$ for every combination of $f_{s0}$, $\sigma_{cen}$, and $\sigma_{sat}$ by requiring that the overall fraction of late-type galaxies matched the observed fraction of late-type galaxies.  

We also employ the `minimal mixing' modeling constraints described by Equations 20-23 of R09 in order to calculate the model $\omega(\theta)$ of early- and late-type galaxies.  These equations place the constraint that early- and late-type galaxies occupy separate halos, as much as the overall statistics allow, onto the calculation of the model $\omega(\theta)$.  Generally, such a model does not allow galaxies of different type to exist in the same low-mass halos but will allow late-type galaxies to exist as satellites, around early-type central galaxies, in more massive halos.  This is opposed to a `full-mixing' model, which places no constraints and thus assumes no environmental dependence other than mass.

The measured $\omega(\theta)$ for early- (top) and late-type (bottom) galaxies are displayed in the left-hand panels of Figure \ref{fig:z4EL}, plotted against the best-fit HOD model (solid lines; same colour scheme as Figure \ref{fig:w2norm}).  The general results are consistent with previous results (e.g., \citealt{Z05}, R09); the early-type galaxies have larger amplitudes than the late-types, and the shape of $\omega(\theta)$ at intermediate angular scales (between about 0.01$^{\rm o}$ and 0.1$^{\rm o}$) is concave for the early-type galaxies and convex for the late-type galaxies.  

Interestingly, the differences between the $0.25 < z < 0.35$ and $0.3 < z < 0.4$ measurements are not dramatic when the galaxies are split by type (unlike for the full sample.  When split by type, the bias values of the best-fit models for the galaxies in the $0.3 < z < 0.4$ slice are only moderately larger than that of the $0.25 < z < 0.35$ slice.  This suggests that the differences in the full sample are driven by the fact that $f_{late}$ decreases dramatically in the $0.3 < z < 0.4$ sample, causing the small-scale amplitudes to be dramatically larger (and a high satellite fraction found for the best-fit HOD) and the overall bias to be larger.  We may therefore wish to consider only the galaxies with $z < 0.35$ when discussing the overall implications of the evolution we observe.

The goodness of fit is not ideal for any of the samples, as the $\chi^2/{\rm DOF}$ presented in Table \ref{tab:HODel}, are significantly greater than 1 in each case.  We note, however, that in each case the minimum $\chi^2$ are far smaller than what is achievable with a model that allows mixing.  The models have difficulty reproducing the shape of the late-type measurements around where they exhibit an increase in slope ($\sim0.02^{\rm o}$).  In each case, the best-fit model increases in slope at a larger angular scale than the measurement does.  This suggests imperfections in model.  We do not attempt to improve the model, but we do note that the assumption that late-type galaxies are distributed in halos like an NFW profile is probably incorrect (as one would infer from the morphology-density relationship, see, e.g., \citealt{Dre80}).

The shapes of the best-fit HODs of the early- (top) and late-type (bottom) galaxies are displayed in the right-hand panels of Figure \ref{fig:z4EL}.  The $f_{c0}$, $f_{s0}$, $\sigma_{cen}$, and $\sigma_{sat}$ parameters that define these fits are are presented in Tabel \ref{tab:HODel}.  The mass cutoff profiles show similar behavior as the best-fit HODs of the full samples --- the cut-off grows increasingly sharp with redshift for both the early- and late-type best-fit HODs.  The shapes of the late-type HODs show significant differences at around $10^{12} h^{-1}M_{\odot}$, where each HOD shows a local maximum.  Closer inspection reveals that the values of $\sigma_{cut}$, $\sigma_{cen}$, and $f_{c0}$ are strongly correlated --- a smaller $\sigma_{cut}$ results in a smaller $\sigma_{cen}$, a larger $f_{c0}$, and a sharper peak at the local maximum.  We thus do not attribute any special significance to this behaviour.  The evolution we find in the early-/late-type HODs appears to be driven primarily by the evolution of the best-fit HODs of the full samples. 

At larger scales, where the 2-halo term dominates, there is good agreement between the models and the measurements, and we can therefore trust the bias of each best-fit model (presented in Table \ref{tab:HODel}).  The bias increases by $\sim 14\%$ and $\sim 12\%$ for the late- and early-type galaxies, respectively. This increase in bias is significantly greater than the few percent change one would expect of  a passively evolving sample (for $b = 1.43$ at $z =0.34$, the passively evolved sample would have $b = 1.39$ at $z = 0.16$; for $b = 1.16$ at $z =0.34$, the passively evolved bias would be 1.15 at $z = 0.16$).  

The fact that both the early- and the late-type galaxies display significant increases in bias over that of a passively evolving sample means that we cannot attribute the evolution in bias that we observe in the full sample to galaxies of a certain type.  Either the average halo bias of galaxies of both type is decreasing, or the contamination between our samples is large enough (due to, e.g., edge-on late-type galaxies reddened by dust lanes; \citealt{Masters10}) to cause the bias of both samples to decrease.  The main conclusions we can draw from our measurements of the correlation functions of early- and late-type galaxies are that the minimum mixing model continues to be favoured over one that allows uninhibited mixing and that the inconsistencies we found in the clustering of the $0.3 < z < 0.4$ galaxies are due primarily to this sample's relatively low late-type fraction.

\section{Cross-Correlations as a Systematic Test}
One issue with the potential to cause systematic errors in the interpretation of our measurements is our estimation of the redshift distribution of each of our galaxy samples.  If our distributions were systematically affected such that we over-predict the width of the distributions, we would incorrectly interpret our clustering measurements as having a higher bias.  If, for example, the magnitude of this problem grew with redshift, it would cause us to (incorrectly) determine that the bias was growing larger with redshift.  

One way to test the accuracies of our redshift distributions is to perform an angular cross-correlation measurement, $\omega_x(\theta)$, between redshift bins.  We therefore measure three $\omega_x(\theta)$: between $0.1 < z < 0.2$ and $0.2 < z < 0.3$; $0.15 < z < 0.25$ and $0.25 < z < 0.35$; and $0.2 < z < 0.3$ and $0.3 < z < 0.4$.  In each of these cases, any cross-correlation signal is due to the fact that, while there is no overlap in the photometric redshift, the errors on the photometric redshifts imply that the true redshift distributions overlap (as can be seen clearly in Figure \ref{fig:nzDR7}).  One can define a factor $W_x$ as
\begin{equation}
W_x = 1/c \int_{0}^{\infty}{\rm d}z ~H(z)dN/dz_1dN/dz_2 ,
\label{eq:wx}
\end{equation}
where we have simply replaced $(dN/dz)^2$ from Equation (\ref{eq:wnz}) with the multiple of the two redshift distributions of the galaxies that are being cross-correlated.  

Similarly to $W$, the $W_x$ factor quantifies how much of the underlying clustering signal we should observe.  We can expect that this underlying clustering signal of these cross-correlations should be similar to the real-space clustering of the galaxies in the intervening bin (i.e., for the cross-correlation between the $0.1 < z < 0.2$ and $0.2 < z < 0.3$ redshift slices, the real-space galaxy clustering signal should be similar to that of the $0.15 < z < 0.25$ sample).  Therefore, we can expect, based on our redshift distributions, to measure a cross-correlation signal with amplitude $W_x/W_i\omega_i(\theta)$, where $W_i$ and $\omega_i(\theta)$ are calculated using the redshift distribution of the intervening redshift slice.  If we do not measure such a signal, it implies that our redshift distributions may be estimated incorrectly.  

\begin{figure}
\includegraphics[width=84mm]{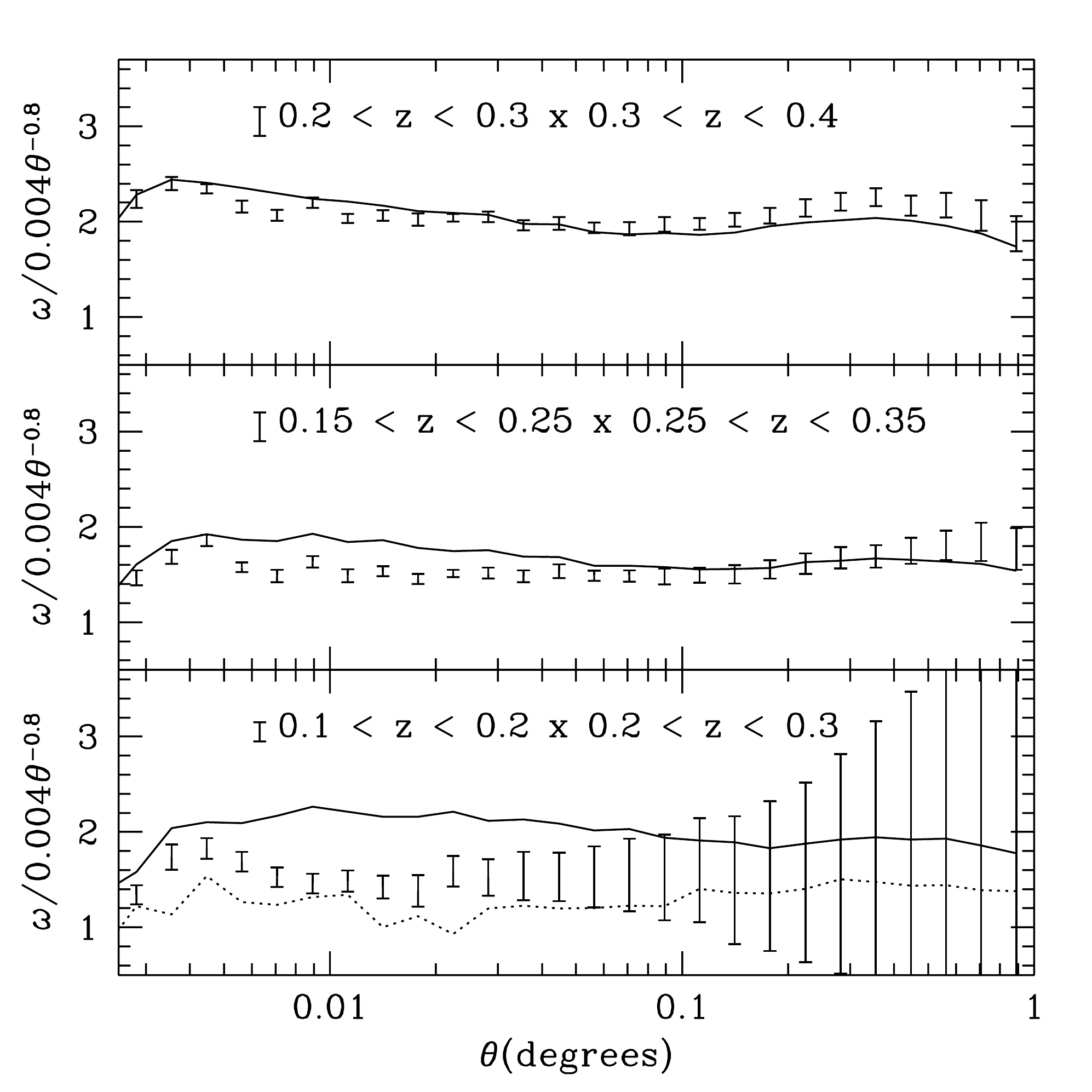}
\caption{The measured (error bars) angular cross-correlation functions between galaxies in the $0.1 < z < 0.2$ and $0.2 < z < 0.3$ redshift slices (bottom), the $0.15 < z < 0.25$ and $0.25 < z < 0.35$ redshift slices (middle), and the $0.2 < z < 0.3$ and $0.3 < z < 0.4$ redshift slices (top), all with amplitudes divided by the power-law 0.004$\theta^{-0.8}$.  In each panel, the measured auto-correlation, multiplied by $W_x/W$ and divided by the same power-law, of galaxies from the intervening bin ($0.15 < z < 0.25$ bottom, $0.2 < z < 0.3$ middle, $0.25 < z < 0.35$ top) is displayed with a solid black line.  In the bottom panel, the dashed line displays the measured auto-correlation of late-type galaxies with $0.15 < z < 0.25$, multiplied by $W_x/W$, again divided by the power law. }
\label{fig:z4cross}
\end{figure}

The three panels of Figure \ref{fig:z4cross} display the three cross-correlations (error-bars) we measure compared to the measured auto-correlation multiplied by $W_x/W$ (solid lines) of the galaxies in each of the respective intervening bins.  The bottom panel displays the cross-correlation of the $0.1 < z < 0.2$ and $0.2 < z < 0.3$ redshift slices (black error-bars), compared to the measured auto-correlation of the $0.15 < z < 0.25$ redshift slice multiplied by $W/W_x$ (solid black line).  All of the data displayed is divided by the power-law 0.004$\theta^{-0.8}$, for clarity.  At large scales, where the 2-halo term dominates, the measurements are consistent.  The error-bars on the cross-correlation are quite large.  This is due to the fact that the overlap between the two redshift slices is small ($W_x/W$ is 0.278).

The overlap between the $0.1 < z < 0.2$ and $0.2 < z < 0.3$ redshift distributions is mainly at the high-redshift tail of the $0.1 < z < 0.2$ slice and the low-redshift tail of the $0.2 < z < 0.3$ slice (see Figure \ref{fig:nzDR7}).  This suggests that much of the cross-correlation signal is due to pairs of relatively high luminosity late-type galaxies (from the $0.1 < z < 0.2$ slice) and relatively low luminosity late-type galaxies (from the $0.2 < z < 0.3$ slice), since late-type galaxies have the highest redshift errors and are thus most likely to occupy the tails of the redshift distribution.  The dashed line in the bottom panel of Figure \ref{fig:z4cross} displays the measured autocorrelation of late-type galaxies with $0.15 < z < 0.25$, multiplied by $W_x/W$.  Its values are more consistent with the measured cross-correlation, suggesting that late-type galaxies do indeed dominate the clustering signal of cross-correlation.  The late-type amplitudes are smaller than that of the cross-correlation, suggesting that while the late-type galaxies dominate measurement, there is still some contribution from early-type galaxies (as one would expect since they are included in the measurement).  We therefore believe this cross-correlation measurement is consistent with the redshift distributions we estimate for the $0.1 < z < 0.2$, $0.15 < z < 0.25$, and $0.2 < z < 0.3$ slices

The middle panel of Figure \ref{fig:z4cross} displays the cross-correlation between the $0.15 < z < 0.25$ and $0.25 < z < 0.35$ redshift slices (black error-bars) compared to the auto-correlation of the $0.2 < z < 0.3$ redshift slice multiplied by $W_x/W$ (solid black line).  At large scales, the two measurements agree extremely well, and we once again measure smaller amplitudes for the cross-correlation at small scales, but to lessor degree than in the bottom panel.  This makes sense given our previous explanation, as the overlap between these redshift slices is greater ($W_x/W$ has increased to 0.327) and the resulting cross-correlation is not dominated to the same degree by galaxies that are likely to be found in the tails of the redshift distribution.  Thus, once more we find the cross-correlation measurement to be consistent with our estimation of the redshift distributions, in this case those for the $0.15 < z < 0.25$, $0.2 < z < 0.3$, and $0.25 < z < 0.35$ slices. 

The top panel of Figure \ref{fig:z4cross} displays the cross-correlation between the $0.2 < z < 0.3$ and $0.3 < z < 0.4$ redshift slices (black error-bars) compared to the auto-correlation of the $0.25 < z < 0.35$ redshift slice multiplied by $W_x/W$ (solid black line).  In this case, the measurements do not agree as well at large scales, but do agree better at small scales.  The measured cross-correlation is larger than we would expect at large scales.  This suggests two possibilities; either the bias of galaxies that contribute to the cross-correlation measurement are significantly higher than those in the $0.25 < z < 0.35$ redshift slice, or $W_x$ is in truth larger than we calculate because we have incorrectly estimated the redshift distributions.  The simplest explanation is that the redshift distribution of the $0.3 < z < 0.4$ is wider than we have estimated.  This would imply only that the bias for this redshift slice has been under-estimated, and it would only strengthen our findings that the bias of galaxies increases with redshift beyond what one would expect from passive evolution.  

Our cross-correlation measurements suggest that our estimation of the redshift distributions has not introduced a systematic error that could be responsible for the trend we find with bias.  We are, therefore, confident that the trend we find --- that with increasing redshift the linear bias of galaxies with $M_r < -21.2$ increases significantly beyond that of a passively evolving sample of galaxies --- is real.  

\section{Physical Interpretation}
\label{sec:phys}
We find that the bias of galaxies with $M_r < -21.2$ increases with redshift significantly beyond the increase in bias one expects of a passively evolving system.  This implies that the galaxies themselves must be evolving under the influence of physical interactions, such as mergers, star formation, or cannibalisation of satellites, etc..  We can rule out the passive effect of ageing stellar populations as a cause for our observations.  This effect would cause the galaxies to dim and therefore suggests that galaxies at $z \sim 0.3$ should be compared to less luminous galaxies at $z \sim 0.1$.  However, if we were to include less luminous galaxies in our $0.1 < z < 0.2$ redshift slice, we would measure an even lower bias.  We therefore know that attempting to account for such evolution, as measurements of the luminosity function (e.g., \citealt{Blan03}) suggest is present, would only enhance the trend we observe.  

If we had cut at approximately $L^*$ for each sample (which would have meant a difference of $\sim$ 0.3 in $M_r$ over our full redshift range based on the evolutionary factors determined by \citealt{Blan03}), the increase in bias with redshift that we observed would have been even stronger.  This implies that our $M_{cut}$ values likely would have been smaller at low redshift than at high.  \cite{ZZ07} found that galaxies from DEEP2 had a higher $M_{cut}$ than those selected from SDSS with similar $L/L^*$ values, which is thus in agreement with our results.  Using a fixed luminosity cut, however, allows us to obtain a fundamental result --- galaxies of the same luminosity reside in halos of approximately the same mass, independent of any change in redshift between 0.1 and 0.4.

\subsection{HOD versus z=0 Bias}
Our best-fit HOD models  imply that the HOD evolves very little with redshift, but this lack of evolution actually implies significant evolution in the large-scale bias of the galaxies.  In order to remove the effect of halo mass growth between redshift slices, we can match HODs by determining the present day bias of passively evolved dark matter halos.  Evolution in bias is not deterministic, but is statistical in nature: the mass of each halo will follow an evolutionary track with a strong stochastic element, that is independent from the large-scale clustering.  When analysing a sample of halos at different redshifts, it is therefore difficult to try to disentangle the effects of structure growth with the growth of galaxies. However, we can use the known bias evolution for a passively evolving population, as a weighted prediction for the evolution of the bias of each sample. 

We adopt the following procedure:  First, we calculate $B(M)$ for each halo mass of the given best-fit HOD (which is evaluated at its own respective median redshift).  We then take this bias and use Equation (\ref{eq:b0}) to obtain the passively evolved $z=0$ bias of the halo, which we denote $b_{\rm o}$.  Comparing our best-fit HODs as a function of this statistic removes the passive effect of large-scale evolution, leaving the effects of mergers and galaxy formation.  The effect of mergers should be small as relative effects on the bias of galaxies leaving a sample, and those joining, should cancel to a large degree.  Thus, for each HOD, we obtain $\langle N\rangle$ as a function of $b_{\rm o}$, giving us HOD results at different redshifts that we can compare, having accounted for halo growth in the mean.

\begin{figure}
\includegraphics[width=84mm]{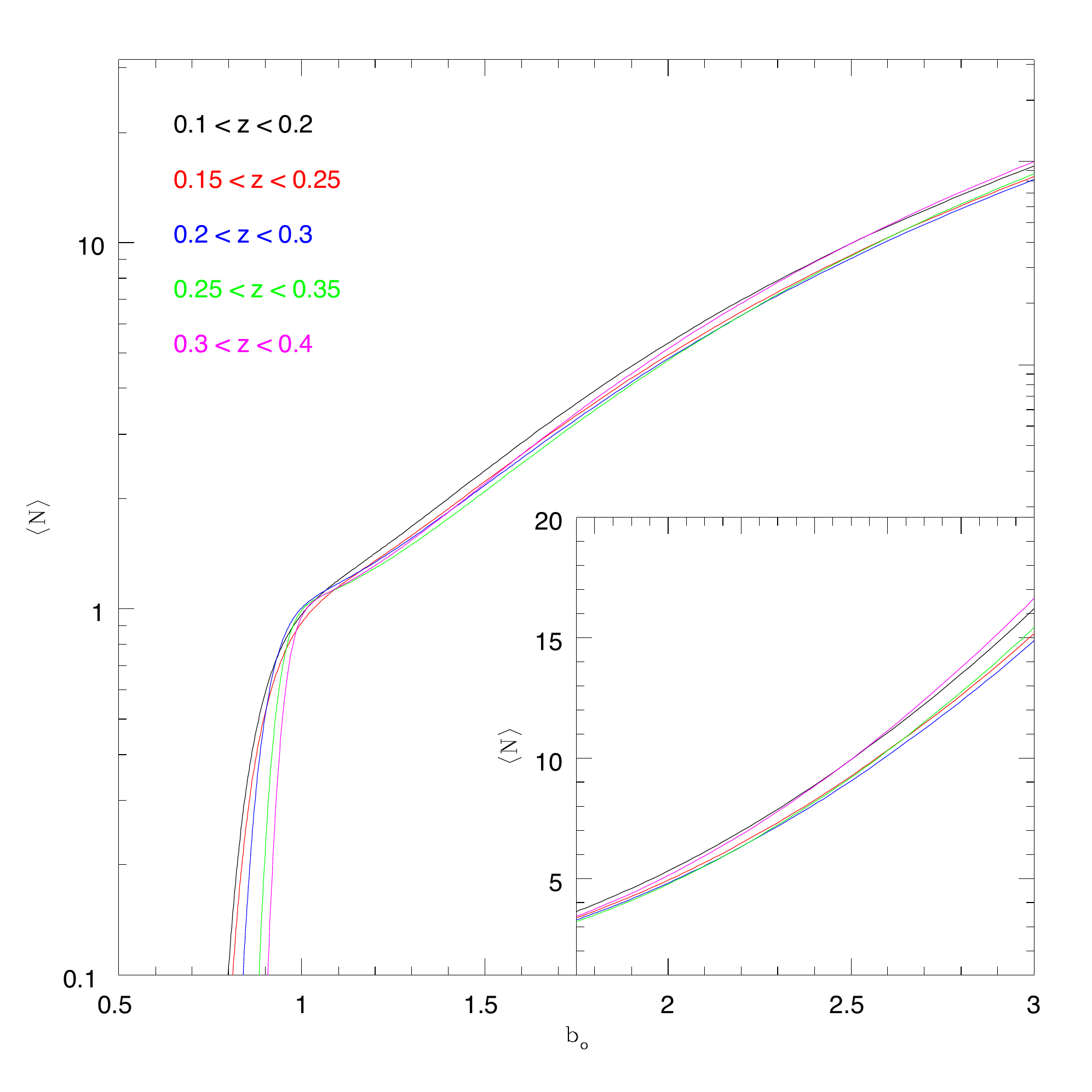}
\caption{The best-fit HOD models for the five photometric redshift slices, plotted against the passively evolved present day value of the halo bias.  The inset plot displays the same data at high $b_o$ with $\langle N\rangle$ scaled linearly.  The colour scheme is the same as for previous plots( $0.1 < z < 0.2$ (black), $0.15 < z < 0.25$ (red), $0.2 < z < 0.2$ (blue), $0.25 < z < 0.35$ (green), $0.3 < z < 0.4$ (magenta)).  }
\label{fig:b0}
\end{figure}

In Figure \ref{fig:b0} we plot (with the same colour scheme as Figure \ref{fig:w2norm}) the mean occupation of galaxies versus $b_{\rm o}$.  The curves at the high-bias end are extremely similar, and there is no trend with redshift.  The major difference between each curve is the value of the bias at which the HOD shows significant decline.  This suggests that the difference between our samples is due almost entirely to evolution in the minimum halo bias required for a halo to host a galaxy with $M_r < -21.2$.  

\begin{figure}
\includegraphics[width=84mm]{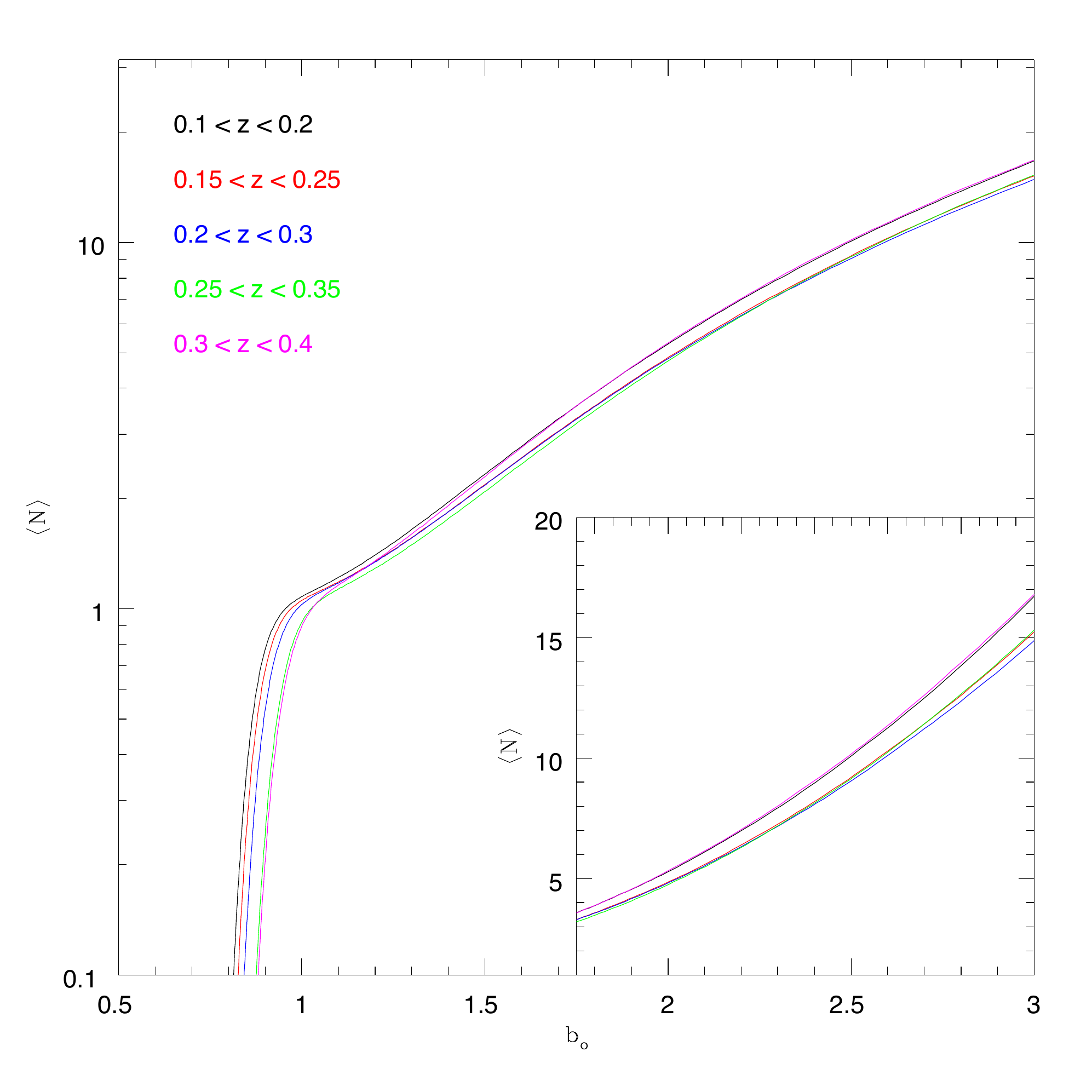}
\caption{Same as Figure \ref{fig:b0}, only for $\sigma_{cut}=0.3$.}
\label{fig:b0sfix}
\end{figure}

Figure \ref{fig:b0sfix} plots the best-fit HODs against $b_{\rm o}$ for $\sigma_{cut} = 0.3$.  All five curves lie nearly on top of each other at $b_{\rm o} = 1.1$.  Going to lower $b_{\rm o}$ values, they separate such that the cut-off $b_{\rm o}$ decreases with redshift, just like as in Figure \ref{fig:b0}.  Thus, both figures suggest that at low redshift, galaxies with $M_r < -21.2$ exist in halos in which they did not exist in at higher redshift, i.e., $\sim L^*$ galaxies are being created (due to, e.g.,  galaxies merging, accreting satellite galaxies in low-mass halos, or undergoing a burst of star formation) in halos with masses around our nearly constant value of $M_{cut} \sim 1.2\times10^{12}h^{-1}M_{\odot}$.  

At large $b_{\rm o}$, both Figures \ref{fig:b0} and \ref{fig:b0sfix} show all of the best-fit HODs to exhibit the extremely similar behaviour, as for both the maximum difference is only $\sim10\%$ and there is no trend with redshift (the highest and lowest redshift slices are nearly identical, as are the three samples in the middle).  Our measurements thus require little evolution in the occupation of halos with bias greater than 1.2, as our nearly constant satellite fractions suggest.  Thus, even as they grow in mass, halos with $b_{\rm o} > 1.2$ do not see a significant increase in the number of galaxies that occupy them.  

\subsection{Robustness Against Changes in Covariance Matrices and Cosmology}
\label{sec:robust}
Our physical interpretation is based on results determined from data with covariance matrices estimated with a jack-knife method and models using one specific set of cosmological parameters.  One may therefore worry that our results may not be robust to changes in the way we estimate covariance matrices or the cosmology we assume in the models.  To learn about the degree to which our results depend on the specific form of the covariance matrices we use, we assume the same percentage error on all measurements and re-determine the best-fit HODs for the $0.1 < z < 0.2$ and $0.25 < z < 0.35$ samples.  Fixing $\sigma_{cut} = 0.3$, we find $\alpha$ changes to 1.12 for the $0.1 < z < 0.2$ sample and 1.104 for the $0.25 < z < 0.35$ sample, while log$_{10}(M_1h/M_{\odot}$) changes to 13.330 and 13.258, respectively.  

\begin{figure}
\includegraphics[width=84mm]{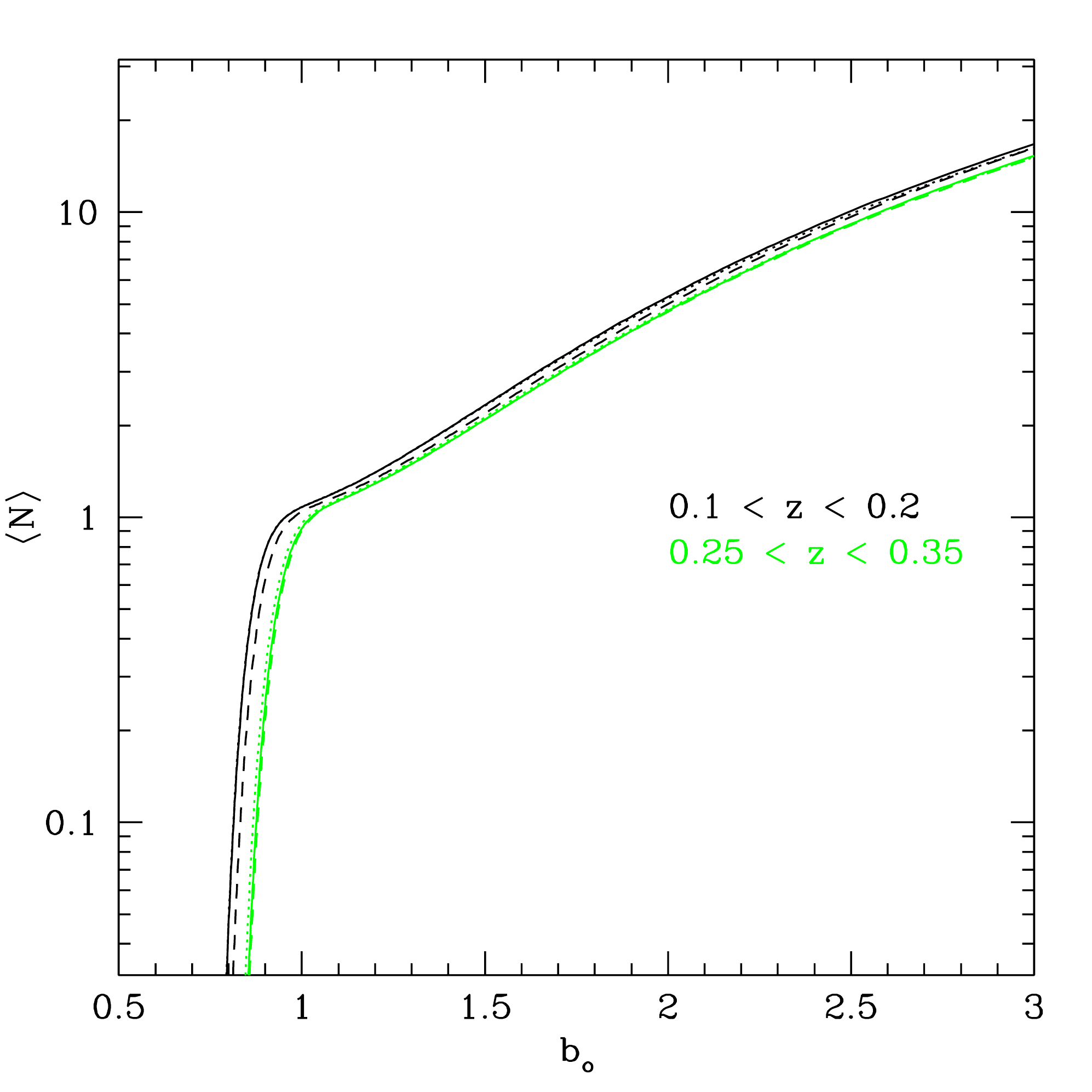}
\caption{The best-fit HODs for $0.1 < z < 0.2$ (black) and $0.25 < z < 35$, with $\sigma_{cut}$ fixed at 0.3, and plotted against the $z = 0$ passively evolved halo bias, $b_{\rm o}$.  The solid lines display the original best-fit (same data as Figure \ref{fig:b0sfix}), the dotted line displays the best-fits obtained when the error on the $\omega(\theta)$ is assumed to be the same percentage in all angular bins, and the dashed line displays the result for a flat universe with $\Omega_m = 0.25$ and $\Omega_b = 0.04$ (and using the original covariance matrices).}
\label{fig:HODbchange}
\end{figure}

The changes in both parameters are minor, as can be seen in Figure \ref{fig:HODbchange}.  This figure displays the best-fit HODs, determined using a fixed percentage error, plotted with dotted lines, and the original data is plotted with solid lines (black for $0.1 < z < 0.2$ and green for $0.25 < z < 0.35$ in both cases), against $b_{\rm o}$.  There are no significant differences (the dotted lines are barely identifiable); clearly treating the error in this manner would have no effect on our physical interpretation.  This suggests that our physical interpretation is robust to any reasonable change in the treatment of our error-bars/covariance matrices, given that the results of \cite{Norberg08} suggest jack-knife covariance matrices should perform far better than simply assuming a constant percentage error.

We also test our results for a different set of cosmological parameters.  We still assume a flat universe, but change $\Omega_m$ to 0.25 and $\Omega_b$ to 0.04.  This change does have a significant effect on the HOD parameters --- $\alpha$ changes to 1.115 for the $0.1 < z < 0.2$ sample and 1.072 for the $0.25 < z < 0.35$ sample, while log$_{10}(M_1h/M_{\odot}$) changes to 13.240 and 13.172, respectively.  The resulting HODs are plotted in Figure \ref{fig:HODbchange} against $b_{\rm o}$ with dashed lines (black for $0.1 < z < 0.2$ and green for $0.25 < z < 0.35$).  While the changes in the HOD parameters are significant, they cause only small changes in the HODs when plotted against $b_{\rm o}$.  For $0.1 < z < 0.2$, the new cosmology causes a slight shift to larger $b_{\rm o}$ values and for $0.25 < z < 0.35$ the two HODs are barely distinguishable.  The lack of change in the best-fit HODs when plotted against $b_{\rm o}$ is due to the fact that the change in the cosmological parameters changes the form of $B(M)$ --- and thus most of the change in the best-fit HOD parameters simply reflects the change in $B(M)$.  Therefore, the changes when the HODs are plotted against $b_{\rm o}$ are minor.  It appears clear that reasonable changes in the cosmology we assume would not cause any significant change in the physical interpretation of our measurements, and our conclusions should therefore be robust.

\subsection{Evolution of Occupation Number}
\begin{figure}
\includegraphics[width=84mm]{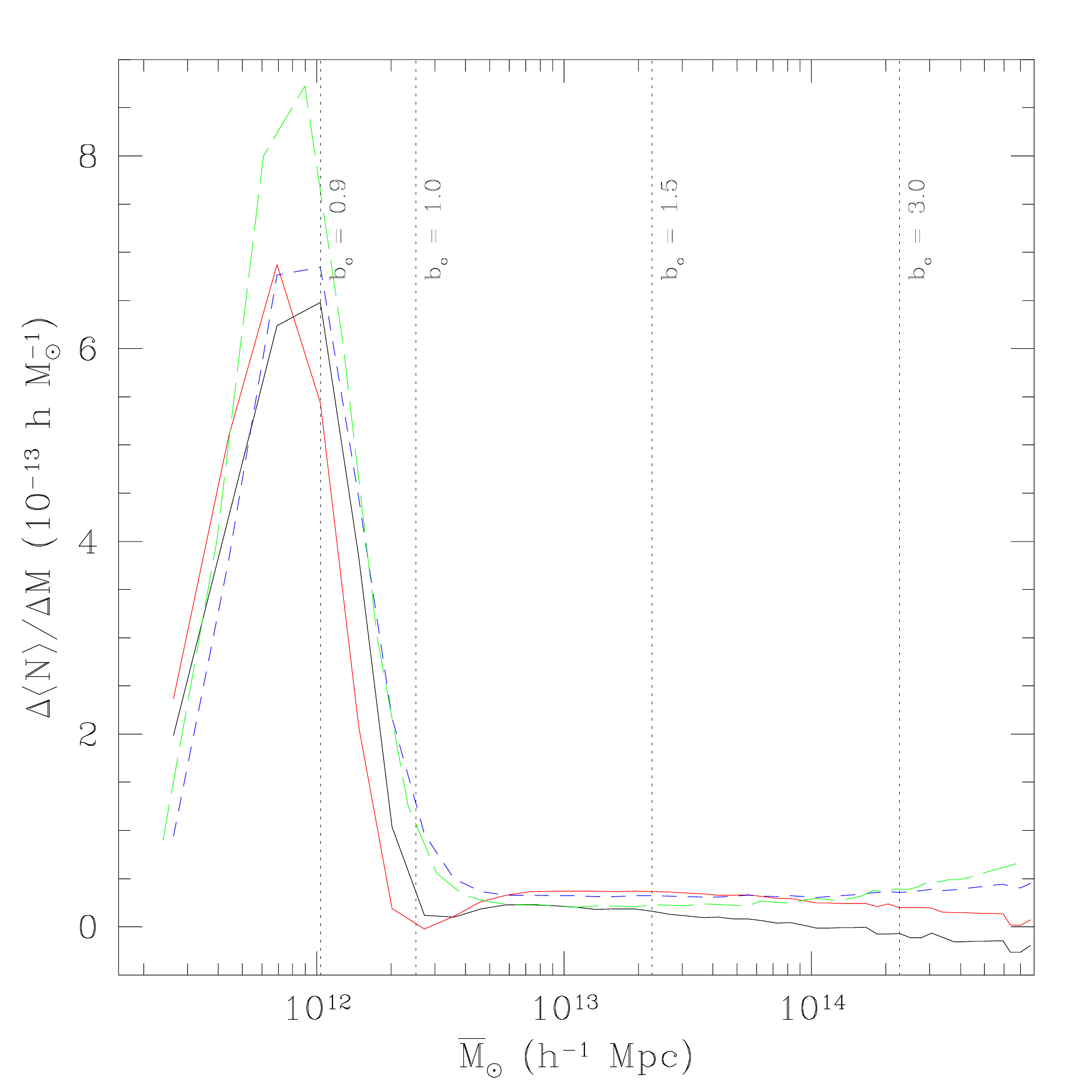}
\caption{The change in occupation number, divided by the change in halo mass, in units of $10^{-13} hM^{-1}_{\odot}$, versus the average halo mass at constant passively evolved $z=0$ bias of the halo.  The solid black line displays result comparing the $0.1 < z < 0.2$ and $0.3 < z < 0.4$ best-fit HODs, the dotted red line displays the result when comparing the $0.1 < z < 0.2$ and $0.25 < z < 0.35$ best-fit HODs, the dashed blue line displays the result when comparing the $0.1 < z < 0.2$ and $0.25 < z < 0.35$ best-fit HODs when $\sigma_{cut}$ is fixed at 0.3, and the long dashed green line displays the result when comparing the $0.1 < z < 0.2$ and $0.25 < z < 0.35$ best-fit HODs when $\sigma_{cut}$ is fixed at 0.3 and the cosmological model is changed to a flat universe with $\Omega_m = 0.25$ and $\Omega_b = 0.04$.  The vertical lines denote selected $b_{\rm o}$ values.}
\label{fig:dndm}
\end{figure}

Another way to look at our results is to plot the change in occupation number divided by the change in mass as a function of $b_{\rm o}$, $\Delta N/\Delta M (b_{\rm o})$.  Figure \ref{fig:dndm} displays this information, in units of $10^{-13} hM^{-1}_{\odot}$, against the {\it average} halo mass for constant $b_{\rm o}$ values (vertical lines denote $b_{\rm o} = 0.9$, 1.0, 1.5, and 3.0 for reference).  The four curves display the change between the $0.3 < z < 0.4$ and $0.1 < z < 0.2$ best-fit HODs (solid black line), the $0.25 < z < 0.35$ and $0.1 < z < 0.2$ best-fit HODs (dotted red line), the $0.25 < z < 0.35$ and $0.1 < z < 0.2$ best-fit HODs with $\sigma_{cut}$ fixed at 0.3 (dashed blue line), the $0.25 < z < 0.35$ and $0.1 < z < 0.2$ best-fit HODs with $\sigma_{cut}$ fixed at 0.3 after changing the assumed (flat) cosmology to $\Omega_m = 0.25$ and $\Omega_b = 0.04$ (long-dashed green line).  

All four curves in Figure \ref{fig:dndm} display a strong peak at $b_{\rm o} \sim 0.9$, and this corresponds to an average halo mass of $\sim 10^{12} h^{-1}M_{\odot}$.  Halos with $b_{\rm o} = 0.9$ gain $\sim 7.5\times10^{11} h^{-1}M_{\odot}$ mass between $z = 0.34$ and $z = 0.16$.  Thus, given $\Delta N/\Delta M$ is $\sim 7 \times 10^{-13} hM^{-1}_{\odot}$, this accreted mass allows the creation of an average of $\sim 0.5$ galaxies per halo.  We have thus presented evidence for the following scenario: the bias of $M_r < -21.2$ galaxies decreases as redshift decreases because these galaxies form preferentially in halos with masses $\sim 10^{12}h^{-1}M_{\odot}$ (independent of redshift) and the bias of these halos naturally decreases as the Universe evolves. 

At the high $b_{\rm o}$ end of Figure \ref{fig:dndm}, our results are inconclusive.  Looking at the evolution between $0.1 < z < 0.2$ and $0.3 < z < 0.4$, our results suggest that highly biased halos are actually losing galaxies, which would imply that the number of galaxies that leave the sample, either through mergers or dimming, is greater than the number of galaxies the halos accrete as they accrete mass.  Our measurements from the $0.3 < z < 0.4$ slice may not be reliable, however, so a comparison between $0.1 < z < 0.2$ and $0.25 < z < 0.35$ (red dotted line) may be more safe.  This comparison suggests that the increase in number slowly decreases, but the same comparison for the best-fit models with $\sigma_{cut} = 0.3$ suggest a slow increase in the number with bias.  In both cases, however, the maximum increase for halos with $\bar{M}_{\odot} > 10^{13}$ is $\sim0.5 \times 10^{-13} hM^{-1}_{\odot}$.  Since all of our best-fit HODs have $\langle N\rangle > 1$ for halo of mass $10^{13} h^{-1}M_{\odot}$, this gain in occupation number does suggest that there is a net loss of galaxies in high mass halos, as would occur if galaxies merge after entering the halo.    

Figure \ref{fig:dndm} suggests a required mass threshold is $\sim10^{12}h^{-1}M_{\odot}$ for $M_r < -21.2$ galaxies.  This result is in basic agreement with numerical models (e.g., \citealt{Shank06}), which show that the fraction of baryonic mass converted into stars peaks at $\sim10^{12}h^{-1}M_{\odot}$ and also the recent results of \cite{Guo}, which suggest galaxy formation efficiency peaks at a halo mass slightly lower than $10^{12}h^{-1}M_{\odot}$.  This implies the following scenario: the ability of a halo to host a galaxy with $M_r < -21.2$ is tied to its ability to convert its baryonic matter into a sufficient number of stars coalesced into a single galaxy, the efficiency with which a halo can do this is related to its mass, and this implies that the bias of the halos in which galaxies form decreases as the Universe evolves.  We therefore find a peak in the $\Delta N / \Delta M$ versus $b_{\rm o}$ relationship at the $b_{\rm o}$ value that corresponds to having the most halos cross the $\sim10^{12}h^{-1}M_{\odot}$ mass threshold.

\section{Conclusion}
\label{sec:con}
We have calculated the angular auto-correlation functions of SDSS DR7 galaxies with $M_r < -21.2$ in five overlapping photometric redshift slices with $0.1 < z < 0.2$, $0.15 < z < 0.25$, $0.2 < z < 0.3$, $0.25 < z < 0.35$, and $0.3 < z < 0.4$ and we found best-fit HODs for each sample by applying the halo model.  The most relevant results are:

\noindent $\bullet$ The bias increases with redshift and the increase is far greater than one would expect of a passively evolving sample of galaxies.

\noindent $\bullet$  When we split our sample by galaxy type into early- and late-type samples, we find the increase in bias is similar for both samples.  We also find that the clustering of early- and late-type galaxies is better fit by a minimal mixing model (as presented in R09) than one that allows galaxies to mix freely within halos, though the high $\chi^2$ values of our best-fit results suggest that the model needs to be further refined. 

\noindent $\bullet$ The best-fit HODs of our full sample suggest that the mass cut-off remains nearly constant with redshift (especially for fixed $\sigma_{cut}$).  Since halos grow in mass as the Universe evolves, one would expect that for a passively evolving sample, the mass cut-off would increase as the redshift gets smaller.  Interpreted in terms of the $z=0$ halo bias, $b_{\rm o}$, this constant mass cut-off implies smaller cut-off value of $b_{\rm o}$ at lower redshift, and this allows the lower redshift galaxies to have a lower overall bias.  This implies that galaxies with $M_r < -21.2$ are forming (via mergers, delayed star formation, accretion of dwarf galaxies, etc.) in increasingly less-biased halos as the Universe evolves.

\noindent $\bullet$  Comparing the change in occupation number versus the change in mass for halos with constant $b_{\rm o}$, we find a strong peak at $b_{\rm o} \sim 0.9$.  This bias value corresponds to an average halo mass of $10^{12} h^{-1}M_{\odot}$, suggesting that galaxies with $M_r \sim -21.2$ form preferentially in halos of mass $10^{12} h^{-1}M_{\odot}$.

\noindent $\bullet$ Our results are consistent with previous measurements made testing the evolution of the HOD (e.g., \citealt{ZZ07}) and numerical models (e.g., \citealt{Shank06}) which predict maximum star formation efficiencies for in halos of mass $10^{12} h^{-1}M_{\odot}$.  

\noindent $\bullet$ Our results are robust against changes in our treatment of the error-bars/covariance and changes in the underlying cosmology.  It would be ideal to confirm our results via simulations or semi-analytic galaxy formation models, but we leave such investigations for future study.

Future surveys will be able to provide significant follow-up to these results.  DES will allow our finding --- that $M_r < -21.2$ galaxies form of constant mass, independent of redshift between $0.1 < z < 0.4$ --- to be tested over a much wider range of redshifts and luminosities.  If our results prove robust, we will be able to determine a fundamental relationship between the luminosity of a galaxy and the mass of the halo in which it forms.

\section*{Acknowledgements}
We thank the referee, Robert Smith, for very helpful comments and suggestions.  We thank David Wake for helpful suggestions.  AJR and RJB acknowledge support for Microsoft Research, the University of Illinois, and NASA through grant NNG06GH156.  The authors made extensive use of the storage and computing facilities at the National Center for Super Computing Applications and thank the technical staff for their assistance in enabling this work.

AJR and WJP thank the UK Science and Technology Facilities Council for financial support.  WJP is also grateful for support from the Leverhulme trust and the European Research Council.

Funding for the creation and distribution of the SDSS Archive has been provided by the Alfred P. Sloan Foundation, the Participating Institutions, the National Aeronautics and Space Administration, the National Science Foundation, the U.S. Department of Energy, the Japanese Monbukagakusho, and the Max Planck Society. The SDSS Web site is http://www.sdss.org/.

The SDSS is managed by the Astrophysical Research Consortium (ARC) for the Participating Institutions. The Participating Institutions are the University of Chicago, Fermilab, the Institute for Advanced Study, the Japan ParticipationGroup, Johns Hopkins University, the Korean Scientist Group, Los Alamos National Laboratory, the Max Planck Institute for Astronomy (MPIA), the Max Planck Institute for Astrophysics (MPA), New Mexico State University, the University of Pittsburgh, the University of Portsmouth, Princeton University, the United States Naval Observatory, and the University of Washington.


\begin{thebibliography}{7}
\bibitem[Abazajian et al.(2009)]{DR7} Abazajian, K.~N., et 
al.\ 2009, ApJS, 182, 543
\bibitem[Baldauf et al.(2009)]{Bald} Baldauf, T., Smith, 
R.~E., Seljak, U., \& Mandelbaum, R.\ 2009, arXiv:0911.4973
\bibitem[Ball et al.(2008)]{Ball08} Ball, N.~M., Loveday, J., \& Brunner, R.~J.\ 2008, MNRAS, 383, 907
\bibitem[Blake et al.(2008)]{Blake} Blake, C., Collister, A., Lahav, O.\ 2008, MNRAS, 385, 1257 
\bibitem[Blanton et al.(2003)]{Blan03} Blanton, M. R. et al.\ 2003, ApJ, 592, 819
\bibitem[Blanton et al.(2005)]{Blanton05sb} Blanton, M.~R., Lupton, 
R.~H., Schlegel, D.~J., Strauss, M.~A., Brinkmann, J., Fukugita, M., 
\& Loveday, J.\ 2005, ApJ, 631, 208
\bibitem[Brown et al.(2008)]{Brown08} Brown, M. J. I. et al.\ 2008, ApJ, 682, 937 
\bibitem[Budav{\'a}ri et al.(2003)]{Bud03} Budav{\'a}ri, T., 
et al.\ 2003, ApJ, 595, 59 
\bibitem[Bullock et al.(2001)]{Bull01} Bullock, J.~S., Kolatt, T.~S., Sigad, Y., Somerville, R.~S., Kravtsov, A.~V., Klypin, A.~A., Primack, J.~R., \& Dekel, A.\ 2001, MNRAS, 321, 559
\bibitem[Cooray \& Sheth(2002)]{CooSh02} Cooray, A., \& Sheth, R.\ 2002, Phys. Rep., 372, 1
\bibitem[Cowie et al.(1996)]{Cow96} Cowie, L.~L., Songaila, A., Hu, E.~M., \& Cohen, J.~G.\ 1996, AJ, 112, 839 
\bibitem[Croton et al.(2006)]{Cr06} Croton, D.~J., Norberg, P., Gazta{\~n}aga, E., \& Baugh, C.~M.\ 2007, MNRAS, 379, 1562  
\bibitem[Dressler(1980)]{Dre80} Dressler,~A., 1980, ApJ, 236, 351
\bibitem[Dressler et al.(1997)]{Dress97} Dressler, A., et al.\ 
1997, ApJ, 490, 577 
\bibitem[Fukugita et al.(1996)]{F} Fukugita, M., 
Ichikawa, T., Gunn, J.~E., Doi, M., Shimasaku, K., \& Schneider, D.~P.\ 
1996, AJ, 111, 1748
\bibitem[Gunn et al.(1998)]{C} Gunn, J.~E., et al.\ 1998, 
AJ, 116, 3040
\bibitem[Guo et al.(2009)]{Guo} Guo, Q., White, S., Li, C., 
\& Boylan-Kolchin, M.\ 2009, arXiv:0909.4305
\bibitem[Jain et al.(2003)]{jain03} Jain, B., 
Scranton, R., \& Sheth, R. K.\ 2003, MNRAS, 345, 62
\bibitem[Jenkins et al.(2001)]{Jenk01} Jenkins, A., Frenk, 
C.~S., White, S.~D.~M., Colberg, J.~M., Cole, S., Evrard, A.~E., Couchman, 
H.~M.~P., \& Yoshida, N.\ 2001, MNRAS, 321, 372 
\bibitem[Kauffmann et al.(1997)]{Kauf97} Kauffmann, G., 
Nusser, A., \& Steinmetz, M.\ 1997, MNRAS, 286, 795
\bibitem[Kravtsov et al.(2004)]{krav04} Kravtsov, A.~V., 
Berlind, A.~A., Wechsler, R.~H., Klypin, A.~A., Gottl{\"o}ber, S., Allgood, 
B., \& Primack, J.~R.\ 2004, ApJ, 609, 35
\bibitem[Landy \& Szalay(1993)]{LZ} Landy S. D., Szalay A. S., 1993, ApJ, 412, 64
\bibitem[Limber(1954)]{lim} Limber, D. N. 1954, ApJ, 119, 655
\bibitem[Lupton et al.(2002)]{L} Lupton, R.~H., Ivezic, Z., Gunn, J.~E., Knapp, G., Strauss, M.~A., \& Yasuda, N.\ 2002, Proc. SPIE, 
4836, 350
\bibitem[Madgwick et al.(2003)]{Ma03} Madgwick, D.~S., et 
al.\ 2003, MNRAS, 344, 847 
\bibitem[Masters et al.(2010)]{Masters10} Masters, K.~L., et al.\ 
2010, arXiv:1001.1744
\bibitem[Mo \& White (1996)]{MW96} Mo, H.~J., White, S.~D., 1996, MNRAS, 282, 347
\bibitem[Montero-Dorta \& Prada(2009)]{DR6LF} Montero-Dorta, A.~D., \& Prada, F.\ 2009, MNRAS, 399, 1106
\bibitem[Myers et al.(2005)]{Mye05} Myers,~A.~D., Outram,~P.~J., Shanks,~T., Boyle,~B.~J., Croom,~S.~M., Loaring,~N.~S., Miller,~L., \& Smith,~R.~J. 2005, MNRAS, 359, 741 
\bibitem[Myers et al.(2006)]{Mye06} Myers, A.~D., et al.\ 
2006, ApJ, 638, 622
\bibitem[Myers et al.(2007)]{Mye07} Myers, A.~D., et al.\ 
2007, ApJ, 658, 85 
\bibitem[Navarro, Frenk, \& White(1997)]{NFW} Navarro J. F., Frenk C. S., White S. D. M., 1997, ApJ, 490, 493
\bibitem[Nishimichi et al.(2006)]{Nish06} Nishimichi, T., Kayo, I., Hikage, C., Yahata, K., Taruya, A., Jing, Y.~P., Sheth, R.~K., \& Suto, Y.\ 2007, PASJ, 59, 93 
\bibitem[Hopkins(2004)]{Hop04} Hopkins, A.~M.\ 2004, ApJ, 
615, 209
\bibitem[Nock et al.(2010)]{Nock} Nock, K., Percival, W.~J., \& Ross, A.~J.\ 2010, arXiv:1003.0896
\bibitem[Norberg et al.(2001)]{Norberg01} Norberg, P., et al.\ 
2001, MNRAS, 328, 64
\bibitem[Norberg et al.(2002)]{N02} Norberg, P., et al.\ 
2002, MNRAS, 332, 827
\bibitem[Norberg et al.(2009)]{Norberg08} Norberg, P., Baugh, 
C.~M., Gazta{\~n}aga, E., \& Croton, D.~J.\ 2009, MNRAS, 396, 19
\bibitem[Padmanabhan et al.(2007)]{Pad07} Padmanabhan, N., 
White, M., \& Eisenstein, D.~J.\ 2007, MNRAS, 376, 1702
\bibitem[Peacock \& Smith(2000)]{PS00} Peacock, J.~A., \& Smith, R.~E.\ 2000, MNRAS, 318, 1144
\bibitem[Peebles(1980)]{P80} Peebles, P.~J.~E.\ 1980, 
Research supported by the National Science Foundation.~Princeton, N.J., 
Princeton University Press, 1980.~435 p.
\bibitem[Ross et al.(2006)]{R06} Ross, A.~J., Brunner, 
R.~J., \& Myers, A.~D.\ 2006, ApJ, 649, 48
\bibitem[Ross et al.(2007)]{R07} Ross, A.~J., Brunner, 
R.~J., \& Myers, A.~D.\ 2007, ApJ, 665, 67 (R07)
\bibitem[Ross \& Brunner (2009)]{R09} Ross, A.~J. \& Brunner, 
R.~J., 2009, MNRAS, 399, 878 (R09)
\bibitem[Schlegel, Finkbeiner \& Davis(1998)]{Sc} Schlegel, D.~J., 
Finkbeiner, D.~P., \& Davis, M.\ 1998, ApJ, 500, 525
\bibitem[Scoccimarro et al.(2001)]{scoc01} Scoccimarro R., Sheth R. K., Hui L., Jain B., 2001, ApJ, 546, 20
\bibitem[Scoville et al. (2006)]{Scoville06} Scoville, N., et al.\ 2007, ApJS, 172, 150
\bibitem[Scranton et al.(2002)]{Scr02} Scranton, R., et al.
2002, ApJ, 579, 48 
\bibitem[Seljak et al.(2005)]{Sel05} Seljak, U., et al.\ 
2005, Phys. Rev. D, 71, 043511
\bibitem[Shankar et al.(2006)]{Shank06} Shankar, F., Lapi, A., 
Salucci, P., De Zotti, G., \& Danese, L.\ 2006, ApJ, 643, 14
\bibitem[Sheth \& Tormen(1999)]{Sheth99} Sheth, R.~K., \& 
Tormen, G.\ 1999, MNRAS, 308, 119
\bibitem[Sheth et al.(2001)]{Sheth01} Sheth, R.~K., Mo, H.~J., 
\& Tormen, G.\ 2001, MNRAS, 323, 1
\bibitem[Skibba \& Sheth(2009)]{skibbasheth09} Skibba, R.~A., \& Sheth, R.~K.\ 2009, MNRAS, 392, 1080  
\bibitem[Smith et al.(2003)]{Smith} Smith, R.~E., et al.\ 
2003, MNRAS, 341, 1311  
\bibitem[Strateva et al.(2001)]{Strat01} Strateva, I., et al. 2001, AJ, 122, 1861
\bibitem[Tinker et al.(2005)]{Tinker} Tinker J. L., Weinberg D. H., Zheng Z., Zehavi I., 2005, ApJ, 631, 41
\bibitem[Tinker et al.(2008)]{Tink08} Tinker, J.~L., Conroy, 
C., Norberg, P., Patiri, S.~G., Weinberg, D.~H., 
\& Warren, M.~S.\ 2008, ApJ, 686, 53
\bibitem[Tojeiro \& Percival(2010)]{To10} Tojeiro, R., \& Percival, W.~J.\ 2010, arXiv:1001.2015
\bibitem[Wake et al.(2008)]{Wake08} Wake D.~A. et al., \ 2008 MNRAS, 387, 1045
\bibitem[White et al.(2007)]{White07} White, M., Zheng, Z., 
Brown, M.~J.~I., Dey, A., \& Jannuzi, B.~T.\ 2007, ApJL, 655, L69
\bibitem[Willmer et al.(1998)]{W98} Willmer, C.~N.~A., da 
Costa, L.~N., \& Pellegrini, P.~S.\ 1998, AJ, 115, 869
\bibitem[York et al.(2000)]{Y} York, D.~G., et al. 2000, 
AJ, 120, 1579 
\bibitem[Yee et al.(2005)]{Yee05} Yee, H.~K.~C., Hsieh, 
B.~C., Lin, H., \& Gladders, M.~D.\ 2005, ApJl, 629, L77 
\bibitem[Zehavi et al.(2004)]{Z04} Zehavi, I., et al.\ 
2004, ApJ, 608, 16
\bibitem[Zehavi et al.(2005)]{Z05} Zehavi, I., et al.\ 
2005, ApJ, 630, 1 
\bibitem[Zheng et al.(2005)]{zheng05} Zheng, Z., et al.\ 2005, ApJ, 633, 791
\bibitem[Zheng et al.(2007)]{ZZ07} Zheng, Z., Coil, A.~L., 
\& Zehavi, I.\ 2007, ApJ, 667, 760
\label{lastpage}
\end{thebibliography}
\end{document}